\documentclass{aa}

\usepackage{graphicx}

\usepackage{txfonts}

\usepackage[colorlinks]{hyperref}
\hypersetup{
     colorlinks=true,
     linkcolor=blue,
     filecolor=blue,
     citecolor =blue,      
     urlcolor=blue,
     }

\newcommand{\T}{$T_{\rm eff}$}
\newcommand{\g}{log($g$)}

\newcommand{\vlos}{{\sl $V_{\rm los}$}}

\newcommand{\vz}{{\sl $V_{\rm Z}$}}

\providecommand{\gaia}{\textit{Gaia}}
\providecommand{\gdr}[1]{\textit{Gaia}~DR{#1}}

\usepackage{amsmath}	
\usepackage{amssymb}
\usepackage{listings}
\usepackage{gensymb}

\begin{document}

   \title{The Great Wave}

   \subtitle{Evidence of a large-scale vertical corrugation \\ propagating outwards in the Galactic disc }

   \author{
E. Poggio \inst{1,2} \and
S. Khanna\inst{1} \and
R. Drimmel\inst{1}  \and
E. Zari\inst{3,4} \and
E. D'Onghia\inst{5,6,1} \and \\
M. G. Lattanzi\inst{1} \and
P. A. Palicio\inst{2} \and
A. Recio-Blanco\inst{2} \and
L. Thulasidharan\inst{5}
 }

   \institute{
   INAF - Osservatorio Astrofisico di Torino, via Osservatorio 20, 10025 Pino Torinese (TO), Italy\\ 
              \email{eloisa.poggio@inaf.it} \and
    Université Côte d’Azur, Observatoire de la Côte d’Azur, CNRS, Laboratoire Lagrange, France 
    \and
    Max-Planck-Institute for Astronomy, K\"{o}nigstuhl 17, 69117 Heidelberg, Germany \and
    Dipartimento di Fisica e Astronomia, Universit\`a di Firenze, Via G. Sansone 1, I-50019 Sesto F.no (Firenze), Italy
    \and
    Department of Astronomy, University of Wisconsin, Madison \and
    Department of Physics, University of Wisconsin, Madison 
    }

\date{Received XXX; accepted XXX}

  \abstract{
  
   We analyse the three-dimensional structure and kinematics of two samples of young stars in the Galactic disc, containing respectively young giants ($\sim$17$\, $000 stars out to heliocentric distances of $\sim$7 kpc) and classical Cepheids ($\sim$3400 stars out to heliocentric distances of $\sim$15 kpc). The vertical structure of the two samples exhibit a consistent shape of the Milky Way's warp, whose amplitude reaches $\sim$700 pc at a Galactocentric radius R $\sim$ 14 kpc. Moreover, both samples show evidence of a large-scale vertical corrugation on top of the warp with a vertical height of $\sim$150-200 pc, extending over a large portion of the Galactic disc between Galactocentric radii $R \sim$ 10-12 kpc in the third Galactic quadrant (galactic longitudes $180^\circ < l < 270^\circ$) and $\sim$12-14 kpc in the second Galactic quadrant ($90^\circ < l < 180^\circ$). Its total length is at least 10 kpc and can possibly reach $\sim$ 20 kpc with the Cepheid sample. The stars in the corrugation exhibit both radial and vertical systematic motions, with Galactocentric radial velocities towards the outer disc of about 10-15 km/s. In the vertical motions, once the warp signature is subtracted, the residuals show a large-scale feature of systematically positive vertical velocities, which is shifted to slightly larger Galactocentric radii with respect to the spatial vertical corrugation (with a phase difference of roughly $\pi/2$), indicating an oscillatory behaviour. A comparison of the observed shift with a simple toy model suggests that the corrugation can be interpreted as a wave propagating towards the outer disc. The wave mapped in this work is located at larger heliocentric distances compared to the Radcliffe wave, a $\sim$2.7 kpc filament of dense gas clouds close to the Sun, and exhibits a larger coverage of the Galactic disc. 
 
   }


   \maketitle


\section{Introduction}
\label{sec:introduction}

Understanding our home Galaxy is one of the greatest challenges in modern astrophysics. In the Galactic disc, where most of the stars are contained, a great variety of physical mechanisms are expected to act in concert. For instance, it has been known for a long time that the disc contains large-scale non-axisymmetric features, including a central bar \citep{Okuda:1977}, spiral arms \citep{Georgelin:1976} and a warp \citep{Burke:1957, Kerr:1957} in the outer parts. Those features can influence the evolution of the Galactic disc, and leave their footprints in the observable properties of the stars. Additionally, vertical and in-plane disturbances in the Galactic disc can be induced by the satellite galaxies that orbit the Milky Way. The co-existence and interplay of both internal (such as the influence of the bar and spiral arms) and external (such as accretion events) physical mechanisms determine how the entire Galactic system evolves with time, and, consequently, the properties of the stars that can be observed today (e.g. their spatial distribution, their motions, the amount of metals in their atmospheres). 

In the era of large astronomical surveys, the physical mechanisms at work in our Galaxy can be investigated in great detail. The unparalleled wealth of data from the \gaia\ satellite \citep{GaiaCollBrown:2016,GaiaCollabBrown:2018,GaiaCollVallenari:2023} and ground-based surveys revealed the richness and complexity of the stellar motions in the Galactic disc, triggering new questions and opening new scenarios. Using astrometry from the first \gaia\ Data Release, the kinematic signature of the warp was revealed by \cite{Schoenrich:2018} as a monotonic increase in \vz{} as a function of angular momentum $L_{\rm Z}$, on top of which an additional "wave-like pattern" was also noted \citep[See also][]{Huang:2018}. Using line-of-sight velocities \vlos\footnote{Here we use the term 'line-of-sight velocity' (\vlos) for the Doppler-shift measured from the spectra and the term 'radial velocity' for the cylindric Galactocentric velocity component ($V_R$).} from \gaia\ Data Release 2 (hereafter \gdr{2}), \citet{Friske:2019} and \citet{Khanna:2019} both noted large scale undulations in $V_{R}$ as a function of angular momentum $L_{\rm Z}$ or Energy. \citet{Poggio:2018} photometrically selected both giants and upper main-sequence stars from \gdr{2} with the help of 2MASS photometry, and mapped over a large extent of the Galactic disc the kinematic signature of the warp in the vertical velocities \vz. \cite{Romero:2019} found a high degree of complexity in the vertical distribution and velocities of OB and Red Giant Branch (RGB) stars, triggering the need for complex kinematic models flexible enough to combine both wave-like patterns and an S-shaped lopsided warp. Not long after this it was shown that the astrometric data for the giants could be modelled with a precessing warp \citep{Poggio:2020, Cheng:2020}. \cite{Chrobakova:2021} also tried to estimate the warp precession rate, but due to their large uncertainties, their result was both consistent with a precessing and a static warp (which they favoured, due to Occam's razor). Using accurate distances and kinematics of classical Cepheids, \cite{Dehnen:2023} and \cite{CabreraGadea:2024} mapped the Galactic disc out to $\sim$ 15 kpc of the Sun, finding not only that the warp is precessing, but also that its precession rate gradually declines with Galactocentric radius. Recently, using stars with measured line-of-sight velocities in \gaia\ Data Release 3 (hereafter DR3), \cite{HrannarMcMillan2024} also found that the warp appears to be rapidly precessing, but additional kinematic disturbances would be needed to match the data.

In the vicinity of the Sun, a Galactic North-South asymmetry was discovered by \citet{Widrow:2012} analysing the number density and bulk velocity of main sequence stars from the SDSS-DR8 \citep[Sloan Digital Sky Survey - Data Release 8,][]{Aihara:2011} and SEGUE \citep[Sloan Extension for Galactic Understanding and Exploration, ][]{Yanny:2009}. Not long after, \citet{Yanny:2013} confirmed and expanded the discovery of \citet{Widrow:2012}, detecting a significant Galactic North-South asymmetry in the number density of K and M dwarf stars from the SDSS-DR9 \citep{Ahn:2012}. Using LAMOST spectroscopic velocities \citep{Cui:2012,Zhao:2012} and PPMXL proper motions \citep{Roeser:2010}, \citet{Carlin:2013} found substructures in bulk velocities of disc stars near the Galactic anticenter. \citet{Williams:2013} analyzed the vertical velocity field of red clump giant stars in RAVE \citep{Siebert:2011}, and detected a rarefaction/compression pattern, suggestive of density wave-like behaviour.  

The hypothesis that a passing satellite or dark matter subhalo can excite coherent oscillations of the Galactic stellar disc has been extensively studied in the literature. It has been shown that encounters with satellite galaxies can excite bending and breathing modes in the Galactic disc \citep[][]{Toomre:1972,Mathur:1990,Weinberg:1991}. Using numerical simulations, \citet{Minchev:2009} showed that an initial energy kick approximating a massive Galactic merger can induce waves in the stellar velocity distribution of an axisymmetric galactic disc. \citet{Gomez:2013} found that a satellite similar to the Sagittarius dwarf galaxy (Sgr) can produce North-South asymmetries and vertical wave-like behaviour in the Galactic disc that qualitatively agrees with the observational results from \citet{Widrow:2012}. Toy-model calculations and simulations of disc-satellite interactions from \citet{Widrow:2014} showed that the response of the disc depends on the relative velocity of the satellite. Using cosmological simulations, \cite{Gomez:2017} detected ‘integral sign’ warps and vertical waves in the stellar discs of Milky Way-sized galaxies.

The \gaia\ phase spiral, discovered by \cite{Antoja:2018}, and later studied by several works \citep[e.g.][]{Binney:2018,Bland-Hawthorn:2019, Laporte:2019,BlandHawthorn:2021, Khoperskov:2019,Hunt:2022,Grand:2023,Alinder:2024,Frankel:2023,Frankel:2024,Tremaine:2023}, suggests that the Galactic disc is somewhat out-of-equilibrium, and is still recovering from a recent perturbation. Several studies indicate that the phase spiral can be generated by a perturbation induced by the passage of the Sagittarius dwarf galaxy \citep{Binney:2018, Bland-Hawthorn:2019, BlandHawthorn:2021}, and this is currently the generally favoured explanation. However, \citet{Bennett:2022} were unable to reproduce the amplitude and wavelength of the vertical perturbation observed in the Milky Way disc testing a range of possible models for Sgr, triggering the need for a more complex solution. Mapping the phase spiral with Gaia DR3, \citet{Hunt:2022} discovered a transition from one-armed ‘bending mode’ spirals in the Solar neighbourhood to two-armed ‘breathing spirals’ in the inner Galaxy, containing signatures of multiple
perturbations.

In the anticenter region, an apparent ring of stars was discovered at low galactic latitudes at 18 kpc from the Galactic center \citep{Newberg:2002,Yanny:2003}. This feature has been called Monoceros Ring, Monoceros Stream, or Galactic Anticenter Stellar Stream \citep{Crane:2003,RochaPinto:2003}. Using data from the Sloan Digital Sky Survey between Galactic longitudes $110\degree <l< 229\degree$, \cite{Xu:2015} detected an asymmetry in the number counts of main-sequence stars, which they explained as roughly concentric disc oscillations, opening in the direction of the Milky Way's spiral arms. On the other hand, using the Pan-STARRS1 survey, \cite{Morganson:2016} mapped the three-dimensional structure of the Monoceros ring, finding that it is composed of two roughly concentric arcs, which do not appear to align with the Milky Way spiral arms. Further from the Galactic center than the Monoceros Ring, additional substructures have been detected, like the Triangulum Andromeda Stream \citep{Majewski:2004,RochaPinto:2003, Martin:2007, Price-Whelan:2015}, or the Pisces globular cluster stream \citep{Bonaca:2012,Martin:2014}. Vertical corrugations have also been observed in external galaxies \citep[e.g.][]{Matthews:2008a,Matthews:2008b}.

Using high-resolution N-body simulations, \cite{BlandHawthorn:2021} found that a satellite galaxy similar to Sgr can generate a corrugated bending wave. \cite{TepperGarcia:2022} found that the satellite-induced corrugation is present both in the stars and gas; the corrugation appears to be initially in phase, the two components move apart after a few rotation periods (500-700 Myr).

In this contribution, we explore the vertical structure and kinematics of the Galactic disc, looking for vertical perturbations on top of the warp of the Milky Way. We use two samples of young stellar populations: (i) a sample of young giant stars and (ii) the Cepheids catalog by \citet[][S24 hereafter]{Skowron:2024}. Specific details on our datasets and on their selection are given in Section \ref{sec:data}, as well as a description of the derivation of their distances. In Section \ref{sec:overview}, we give an overview of the young giant sample. For an overview of the Cepheids sample, we refer the reader to the catalog paper S24, as well as a recent application of their data to map the large-scale spiral structure of the Milky Way \citep[][]{Drimmel:2024}. We analyse the vertical distribution of our two samples in Section \ref{sec:analysis}, and their kinematics in Section \ref{sec:kinematics}. We discuss the obtained results in Section \ref{sec:discussion}, and finally present our Summary $\&$ Conclusions in Section \ref{sec:summary_conclusions}.

\section{Data}
\label{sec:data}

This work is based on two different datasets, which are described in the following. 

   \noindent \paragraph{Young giant sample}
   This dataset has been selected from the \gdr{3} catalog \citep{GaiaCollVallenari:2023} with the specific objective to study young populations in the Galactic disc using both spatial and kinematic information. As illustrated in detail below (see Section \ref{subsec:sample_selection}), we want to select stars with typical effective temperatures \T\, $\sim 4500 - 5000$ K and surface gravities \g\, of about $\sim$ 1 dex. This corresponds to the blue loop evolutionary stage \citep[see the massive sample in ][]{DR3-DPACP-104}, and, specifically, to its cold phases \cite[see Sample A in][]{Poggio:2022}. These tracers have been found to map the segments of the nearest spiral arms in the Galaxy, as expected for young populations in the Galactic disc. Indeed, a comparison between theoretical isochrones in the range of typical disc metallicities indicates that this portion of the \T - \g\, diagram is expected to be populated by young stars, approximately less than 100 Myr old (see Figure \ref{fig:selection_kiel_diagram}, more details in Section \ref{subsec:sample_selection}). Those stars are also expected to be typically bright (with absolute magnitudes between -3 and -7 magnitudes in the G band), making them visible out to relatively large distances. To select our sample, we first perform a preliminary selection based on the astrophysical parameters from \citet{Andrae:2023} (described in Section \ref{subsec:sample_selection}); then we infer Bayesian distances to each star using a prior specifically constructed to be self-consistent with the adopted selection criteria, and compare the obtained distances with other values found in the literature (Section \ref{subsec:distances}); finally, we refine our selection with additional cleaning criteria (Section \ref{subsec:contaminants}). We end up with a catalog of 17030 young giant stars, which sample the Galactic disc out to approximately 7 kpc in heliocentric distance. \gdr{3} line-of-sight velocities are available for 14219 young giant stars, which is the great majority of the sample ($\simeq$ 83\%).  \break
   
   \noindent \paragraph{Cepheid sample.} This dataset is based on the classical Cepheids catalog recently presented by S24, which contains new distances based on mid-infrared photometry for 3425 stars, out to approximately 15 kpc in heliocentric distance. From their catalog, we use the 'd\_av' and 'dd\_av' columns as the heliocentric distance and uncertainty, respectively. Typically, the distance uncertainties are less than 5\%, however S24 suggest the uncertainties could be up to 13\% at most. Following \cite{Drimmel:2024}, we remove the stars whose distances are clearly inconsistent with their astrometry, applying a cut Q<5 (where the Q parameter is defined in Equation 14 of S24). After applying this cut, we obtain 3357 stars, of which 2034 also have available line-of-sight velocities and proper motions from the \gdr{3} catalog, allowing us to calculate the full three-dimensional kinematic information for each star. For additional details on the Cepheids sample, we refer the reader to \cite{Skowron:2024} and \citet{Drimmel:2024}.  \break

Our two datasets are clearly complementary: the Cepheid sample reaches very large distances, but with only three thousand stars; on the other hand, the young giant sample contains approximately 5 times the number of stars in a relatively smaller region (out to a distance of $\sim$ 6-7 kpc), therefore with a denser coverage. It should be noted, however, that our two samples are both young, and therefore are expected to trace similar substructures, providing two different views of the young Galactic disc, from independent subsamples of the same young stellar population.

\subsection{Preliminary selection of the young giant sample}
\label{subsec:sample_selection}

\begin{figure}
\centering
\includegraphics[width=0.5\textwidth]{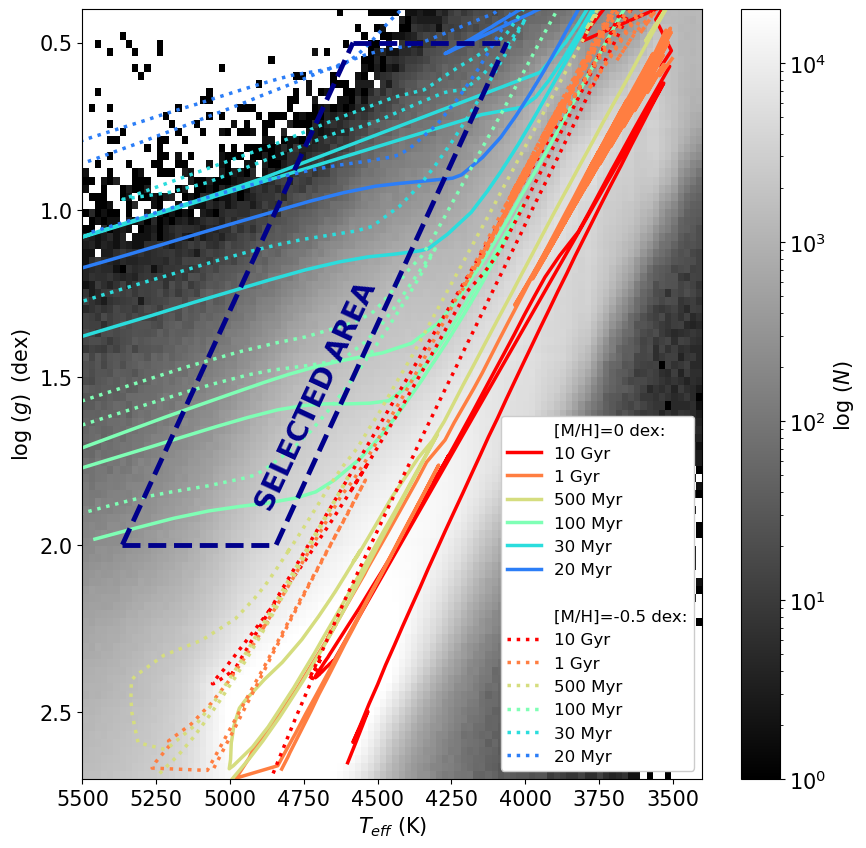}
\caption{ Comparison between the distribution of XGBoost stars \citep{Andrae:2023} in the Kiel diagram and the PARSEC isochrones for different metallicities. The portion of the Kiel diagram selected in this work is shown by a dark blue dashed line.\label{fig:selection_kiel_diagram}
}
\end{figure}

At the time of writing, the largest publicly available catalog of astrophysical parameters is by \citet{Andrae:2023}, who derive stellar parameters (\T,\g,[M/H]) using \gdr{3} XP spectra \citep{DeAngeli:2023, Montegriffo:2023}, combined with CatWISE photometry \citep{Marocco:2021}, for a total of $\sim$ 175 million stars. We shall refer to it hereafter as the XGBoost sample, after the algorithm they employed. To select young giants in \gaia\ Data Release 3, we take the sources that fall in the region of the Kiel diagram with:
\begin{equation}
\begin{aligned}
&2 <  \text{\g} /  \text{dex}  < 0.5 \\ 
&\text{\g} > ( C \cdot \text{\T} + I_{L} ) \\ 
&\text{\g} < ( C \cdot \text{\T} + I_{R} ) \\ 
\end{aligned}
\end{equation}
where the coefficient C=0.00192 $\rm{ dex \, K}^{-1}$ approximately follows the natural slope of the RGB branch \citep[using the inclination adopted in][ to select their Massive sample, but with a different coverage of the Kiel diagram]{DR3-DPACP-104}, while the intercepts $I_{L} = -8.3$ dex and $I_{R} = -7.3$ determine the position of two inclined lines, respectively at the left and right border of the selected area in Figure \ref{fig:selection_kiel_diagram}. The cut in \g\ helps us to select part of the blue loop stars, keeping only values lower than the red clump, which contains less massive and typically older stars. Such inclined portion of the Kiel diagram follows the selection criteria applied in our previous work \citep{Poggio:2022}, where young giants were selected based on \gaia\ GSP-spec data \citep{RecioBlanco:2023}. In this work, we aim to perform a similar selection, but going to fainter magnitudes, to map a larger portion of the Galactic disc. From the above selected objects, we remove stars having apparent magnitude $G>15$, where we find that the sample starts to become incomplete. Finally, we remove possible globular cluster members using the catalog of \cite{Hunt:2023}. After these cuts, we obtain $138\, 829$ stars.

\subsection{Distances of the young giants sample}
\label{subsec:distances}

\begin{figure*}
\centering
\includegraphics[width=18.cm]{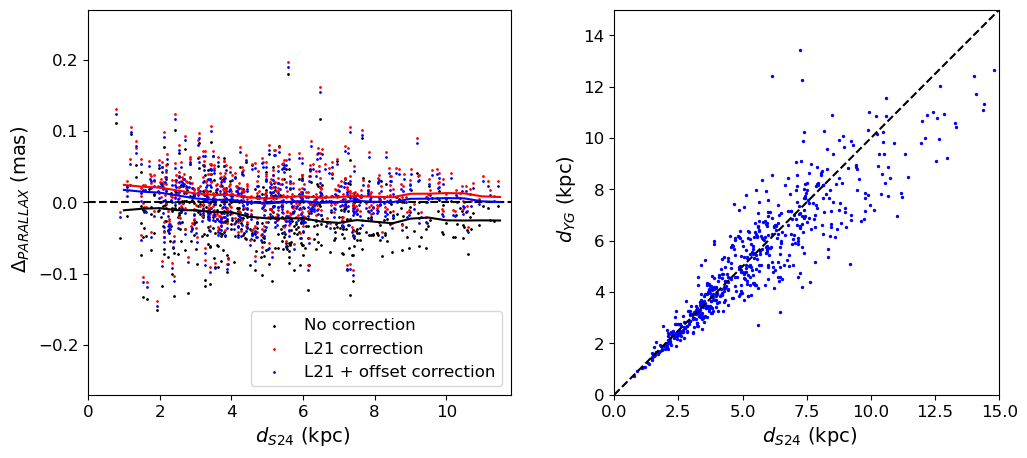}
\caption{ \emph{Left panel}: Comparison between the photometric parallaxes (obtained from Cepheids distance estimates from \cite{Skowron:2024}) and trigonometric parallaxes without zero-point correction (black), with \cite{Lindegren:2021} correction (red), and \cite{Lindegren:2021} + additional offset correction (blue). See main text for detailed explanation. \emph{Right panel}: The derived distances for the 593 young giants ($d_{YG}$) that also are classified as Cepheids, compared to their photometric (Leavitt Law) based distances ($d_{S24}$). \label{fig:comparison_YG_Cepheids}}
\end{figure*}

For each star selected in Section \ref{subsec:sample_selection}, we infer the probability density function (PDF) of the heliocentric distance $d$ for the star's galactic coordinates ($l$, $b$) via Bayes’ theorem, $P (d \, |\,l, b, \varpi_{corr}, \sigma_{\varpi} ) \propto P (\varpi_{corr}\, |\, d, \sigma_{\varpi} ) P (d\, |\, l, b)$, where $\varpi_{corr}$ is the observed parallax corrected for the parallax zero-point offset (see below) and $\sigma_{\varpi}$ the parallax uncertainty. We assume a Gaussian likelihood $ P (\varpi\, |\, d, \sigma_{\varpi} )$, and a prior similar to \citet{Astraatmadja:2016} (see their Equation 7, Milky Way prior):
\begin{equation}
P(d\, |\, l,b) \propto \rho(l,b,d)\, \mathcal{F}(l,b,d) \quad,
\end{equation}
where $\rho(l,b,d)$ is the density distribution of stars in the Galactic disc, and $ \mathcal{F}(l,b,d)$ is the fraction of observable objects (based on the selection function of the catalog, as explained below). Here we assume a simple model for the disc density $\rho(l,b,d)$, composed of an exponential disc in Galactocentric radius $R$, with radial scale length $L_R$ = 2.6 kpc \citep{BlandHawthorn:2016}, and in vertical height Z, with vertical scale height $h_z$ = 150 pc, following \citet{Poggio:2018}. The assumed distance of the Sun from the Galactic center is $R_{\odot}=8.277$ kpc \cite{Gravity:2022}. It is important to note that the simple density prior assumed here does not include non-axisymmetric features (e.g. warp, the spiral arms). This is because non-axisymmetric structures represent one of the main features of interest of this study, and we want to avoid introducing artefacts due to our assumptions. Therefore, any deviation from symmetry observed in the three-dimensional distribution of stars is not expected to be attributable to the assumed prior. The impact of various assumptions incorporated in our prior is discussed in the following sections.

The term $ \mathcal{F}(l,b,d)$ is calculated following the approach of \citet{Astraatmadja:2016} and \citet{Poggio:2018}, but now re-adapted to the specific case of the young giants selected in this work. First, we construct the luminosity function $\Phi(M_G)$ in the G band by applying the same cuts specified in Section \ref{subsec:sample_selection} to the Kiel diagram based on the PARSEC isochrones \citep{Bressan:2012,Chen:2014,Chen:2015,Pastorelli:2019}, assuming the canonical two-part power law IMF corrected for unresolved binaries \cite{Kroupa:2001,Kroupa:2002}, a star formation rate constant with time (which might be realistic over the typical age of the target stars, $\sim$ 100 Myr), and solar metallicity. 

The fraction of observable objects $\mathcal{F}$ is then estimated by integrating the luminosity function for all possible distances along the line-of-sight (after including the completeness of the sample at each apparent magnitude G, see below), down to the faintest observable apparent magnitude \citep[see Equations 4 to 6 in][]{Astraatmadja:2016}, after taking extinction into account. Following the procedure laid out in \cite{Khanna:2024}, we estimate the extinction by combining publicly available dust maps \citep[][etc]{Schlegel1998,Green:2019,Lallement:2019}. Then,
\begin{equation}
\mathcal{F}(l,b,d) = \int_{G_{min}}^{G_{max}} S(l,b,G)\, \Phi(M_G) dG
\end{equation} 
where $S(l,b,G)$ is the selection function of our selected sample and $M_G = G - 5 \log(d) + 5 + A_G(l,b,d)$, for $d$ in parsecs.
To estimate $S(l,b,G)$, we follow the approach laid out in \citet[][]{Castro-Ginard:2023}, by computing the ratio of stars that end up in the catalogue of \citet{Andrae:2023}, to the entire \gaia\ catalogue ($n$) in bins of $(l,b,G)$, between $4<G<15$. On the \gaia\ archive (see \autoref{sec:subsf}) we compute this ratio in bins of $G$ (1 mag wide), and sky position ($l,b$) tiled at \textit{HEALPix} level 5 \citep{HEALPix2005}. Using these number counts, we then take $S(l,b,G)=\frac{k+1}{n +2}$
as the mean probability for a source to end up in our subsample (with $k$ stars) as proposed by \citet[][]{Castro-Ginard:2023}. The completeness at the edge of the brightness limit ($14<G<15$) for the young giants is shown in \autoref{fig:sf_14_15}.

After cleaning the young giant sample (see Section \ref{subsec:contaminants} below), we find 593 Cepheids are also in this sample. 
This is a consequence of the instability strip partially overlying the area of the Kiel diagram used to define the initial selection of these stars. We take advantage of this overlap by using these common stars to check their derived distances against their photometric Cepheid distances, $d_{S24}$, taken from \cite{Skowron:2024}.
We first make a direct comparison of the measured \gaia\ parallaxes with 
the photometric parallax ($1/d_{S24}$). 
Figure \ref{fig:comparison_YG_Cepheids} 
shows the differences between the measured and photometric parallax with no zeropoint correction to the measured parallaxes (black points) and after applying the zeropoint correction of 
\citep{Lindegren:2021} (red points). 
Corresponding running medians (black and red curves) show substantial improvement after the zeropoint correction is applied, though 
a mean offset of 7 $\mu$as is still present, in agreement with the offset found by S24. Such an offset will have an impact on the inferred distances for distant objects (while it is totally irrelevant for nearby stars), and we therefore apply it on top of \cite{Lindegren:2021}'s predicted zero-point offset $zp_{L21}$, that is, $\varpi_{corr} = \varpi - zp_{L21} - 0.007$ mas, where $\varpi$ is the parallax from the Gaia archive. The blue points in Fig \ref{fig:comparison_YG_Cepheids} (left panel) show the results using the final corrected parallaxes, with a running median of the final parallax difference (blue curve) tending to coincide with 0. The right panel of Figure \ref{fig:comparison_YG_Cepheids} shows the comparison between the distances calculated as prescribed above with the photometric distances $d_{S24}$.  

To test the reliability of the distances obtained in this Section, we also cross-match our young giants sample with other catalogs available in the literature \citep[e.g.][]{Hunt:2023,BailerJones:2021}. Within 8 kpc in heliocentric distance (where our sample has the best coverage), 93\% or more of the stars deviate less than 20\% from the distance estimates taken from the literature. We also note that 80\% of the young giant stars have measured parallax $\varpi$ with relative uncertainty less than 20\%, i.e. satisfying $\varpi / \sigma_{\varpi} > 5$. Finally, 91\% of the stars in our catalog have a relative distance error $\sigma_d / d$ < 20\%, where $\sigma_d$ is the average between $\sigma_{up}=d_{84} - d_{50}$ and $\sigma_{down} = d_{50}-d_{16}$, where $d_{84},d_{50},d_{16}$ are, respectively, the 84$^{th}$, 50$^{th}$ and 16$^{th}$ percentiles of the posterior distribution of the distance for each single star. More details are given in Appendix \ref{appendix_sample}.

\begin{figure}
\centering   \includegraphics[width=0.5\textwidth]{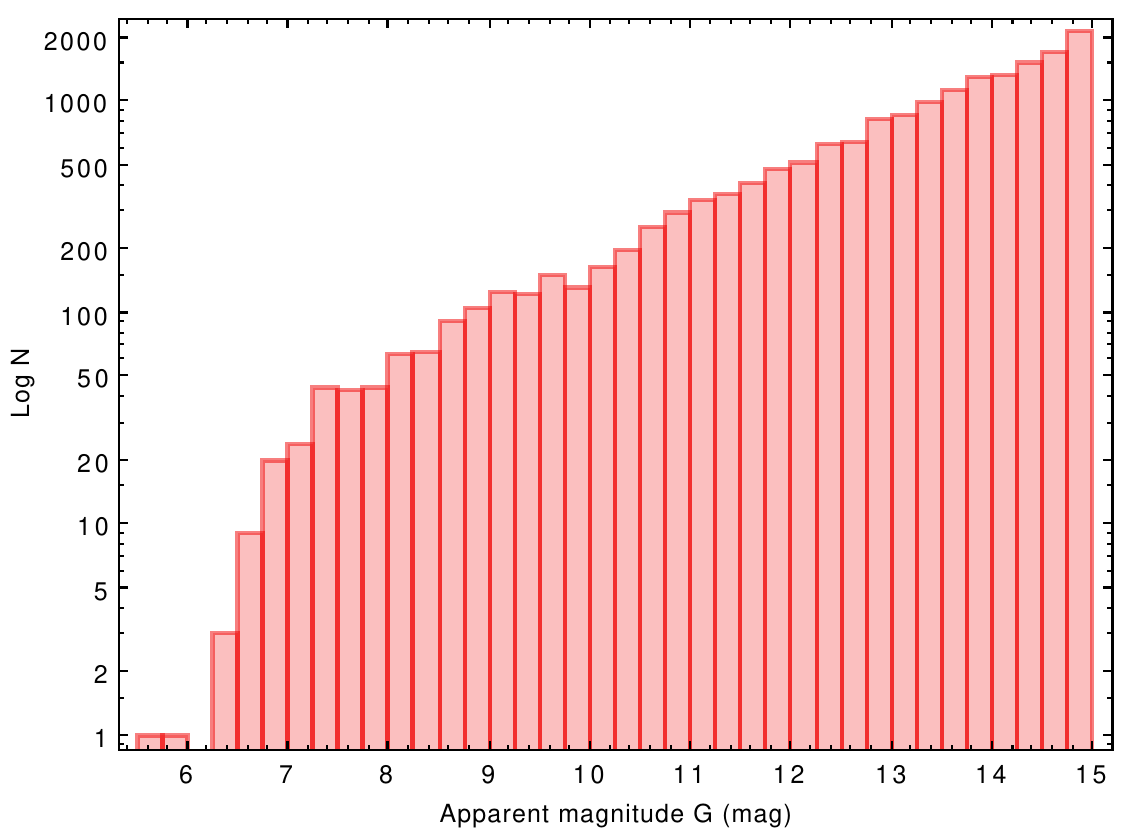}
\caption{ Apparent $G$ magnitude distribution of the young giant sample. \label{fig:Gmag_hist}
}
\end{figure}

\subsection{Cleaning of the young giant sample}
\label{subsec:contaminants}

The selection described in Section \ref{subsec:sample_selection}, which is based on \T\, - \g\, parameters, is not expected to lead to a perfectly pure sample of young giant stars. This is because, in this region of the Kiel diagram, the isochrones of a given age tend to move to the left (i.e. larger \T\ values) for lower values of metallicity, as shown in Fig. \ref{fig:selection_kiel_diagram}. This is a consequence of the decreasing opacity of the atmosphere for stars of lower metallicity. Therefore, it is possible that our selected box in the Kiel diagram does not only contain young stars of disc-like metallicity, but also some metal-poor contaminants of significantly larger age. Additionally, we note that random errors on \T\, - \g\ can make a fraction of the Red Giant Branch stars potential contaminants that fall into the selected region. \citet{Andrae:2023} do not provide individual uncertainties for the single sources, but report the median uncertainties of the sample, i.e. $\sim$ 50 K in \T\, and 0.08 dex in \g, respectively. 

To clean our sample, we remove stars whose metallicity is lower than $[M/H]_{lim} = -0.04 \cdot R - 1.25$, where $R$ is the Galactocentric radius and $-0.04$ dex/kpc follows the typical values of the metallicity gradient of the Galactic disc \citep[e.g.][]{DR3-DPACP-75}. The intercept of our $[M/H]_{lim}$ cut has been tuned to take into account the metallicity dispersion at a constant $R$, and, at the same time, to try to minimize the contamination from metal-poor stars (see Figure \ref{fig:contaminants_RZ_colmet} in Appendix \ref{appendix_sample}). In the inner parts of the Galaxy, we see a significant contribution from metal-poor stars with high velocity dispersions, and therefore we apply a cut at $R <5 \rm{\, kpc}$. Since the above-mentioned metallicity cut is very broad, we refine it by removing the low-angular momentum stars with $V_{\phi} < 180$ km/s \cite[assuming, for the solar velocity, the values reported in ][]{Drimmel:2018}, for the stars that have available radial velocities (> 90\% of the sample). Inside $R<R_{\odot}$, we remove the stars that have both $[M/H]<-1$ dex and $|z| > 0.5$ kpc. We also apply a cut $|b|<30^{o}$, which removes most of the Magellanic Clouds; to better remove stars in the Large Margellanic region on the sky, we additionally remove stars with $b<-28^{o}$ between galactic longitudes $275^{o}$ and $285^{o}$. As a test, we also tried to remove open clusters members from \cite{Hunt:2023}, to insure that our maps (e.g. stellar velocity fields, see next Sections) typically mapped field stars, and were not dominated by individual clusters. Since we obtained very similar results by including or not open cluster members, we finally decided to keep them, to have larger statistics. We also removed 6 stars that are expected to be members of the globular cluster NGC 6656, as explained in Appendix \ref{appendix_sample}.

\section{Overview of the young giant sample}
\label{sec:overview}

\begin{figure}
\centering
    \includegraphics[width=0.5\textwidth]{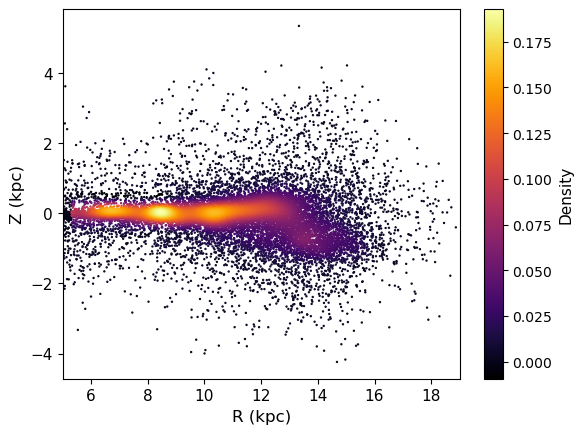}
\caption{ Vertical section of the Galactic disc as a function of Galactocentric radius $R$. Overdense regions are the spiral arms of the Galaxy seen edge-on. Stars are color-coded by the probability density function at each bin, calculated as the number of stars in each bin over total number of stars per bin area, where the bin area is 60 pc (in $R$) times 30 pc (in Z). \label{fig:RZ_allsample}
}
\end{figure}

\begin{figure}
\centering
    \includegraphics[width=0.49\textwidth]{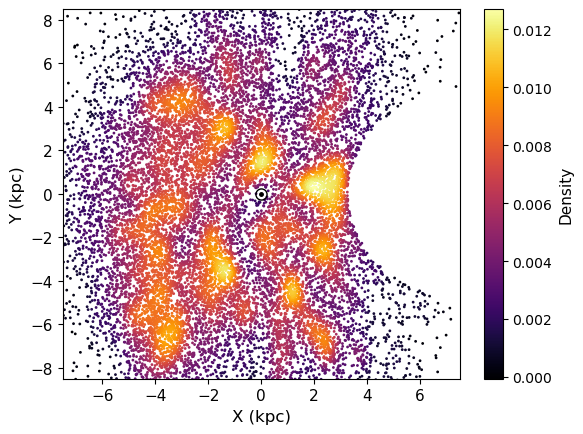}
\caption{ Distribution of the young giant sample in the Galactic plane. The Sun's position is shown by the $\odot$ symbol. The Galactic center is to the right, at (X,Y)=($R_{\odot}$, 0 kpc), with $R_{\odot}=8.277$ kpc \citep{Gravity:2022}. Galactic rotation is clockwise. \label{fig:xy_map_young_giants}
}
\end{figure}

\begin{figure*}
\centering
    \includegraphics[width=0.95\textwidth]{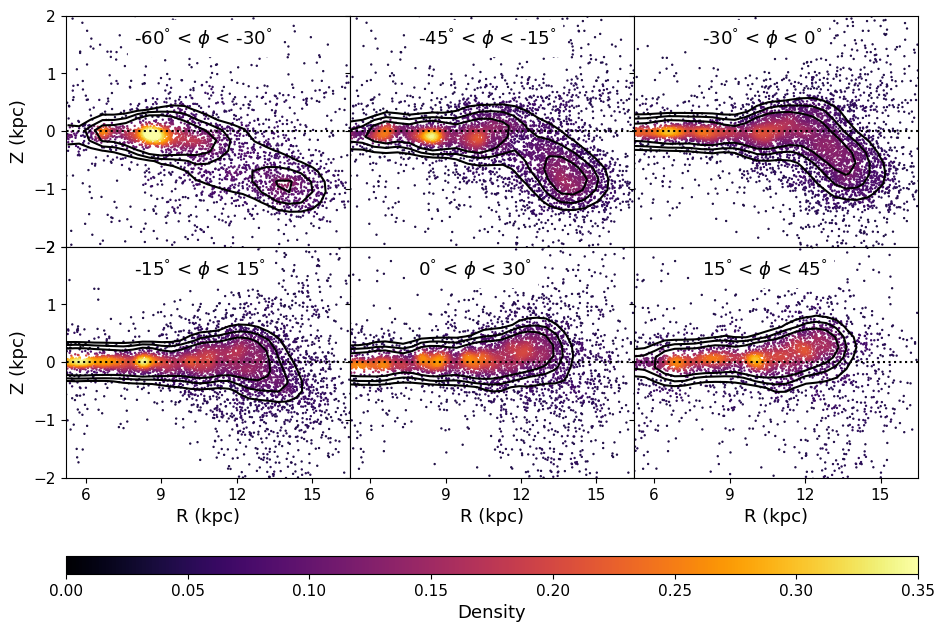}
\caption{ Same as Figure \ref{fig:RZ_allsample}, but now in overlapping bins in Galactic azimuth $\phi$. Azimuthal ranges are chosen to cover the regions where our sample has more data. Isodensity contours are show as black solid lines. \label{fig:overview_warp_flare}
}
\end{figure*}

In this Section, we analyse the three-dimensional distribution of the samples presented in Section \ref{sec:data}. A first overview of the young giants sample is given in Figure \ref{fig:RZ_allsample}, where the distribution of stars in the vertical direction $Z$ as a function of Galactocentric radius $R$ is shown (for stars at all azimuthal angles in the Galactic disc). As we can see, the stars of our sample are located close to the Galactic mid-plane, as expected from young disc stellar populations. However, we note that the distribution of stars in the disc is not continuous: clumps in the stellar density are apparent. Such "blobs" are vertical sections of the Galactic spiral arms. Indeed, the in-plane distribution of the stars shows segments of the nearest spiral arms in the Galaxy, as shown in Figure \ref{fig:xy_map_young_giants}. This is a further indication that our selected stars are expected to be typically young. The position and orientation of the observed segments are in agreement with the spiral structure found by previous works \citep{Zari:2021, Poggio:2021,DR3-DPACP-75}. A more detailed analysis of the in-plane distribution of this sample will be presented in a separate paper.

In the outer regions of the disc ($R\gtrsim 10$ kpc), we can observe a displacement of the Galactic mid-plane with respect to Z=0 kpc, together with an increase of the vertical dispersion as a function of $R$. This is a two-dimensional projection of the Galactic warp and flare, which both act on the outer disc. To have a three-dimensional view of the vertical displacement of the Galactic disc, we now examine Figure \ref{fig:overview_warp_flare}, which is similar to Fig. \ref{fig:RZ_allsample}, but now split in different overlapping slices of Galactic azimuth. (In this work, we define the Galactic azimuth $\phi$ as increasing in the direction of Galactic rotation, and equal to 0$^\circ$ 
toward the Galactic anticenter.)
Figure \ref{fig:overview_warp_flare} shows that the outer disc bends downwards for $\phi \lesssim 0^{\circ}$, and in the opposite direction for positive $\phi \gtrsim 0^{\circ}$. This large-scale distortion of the outer Galactic disc is the well-known Galactic warp, which was first detected in the HI data \cite{Kerr:1957, Burke:1957, Levine:2006}, and later evidenced in different stellar populations in numerous previous works \citep[e.g.][and others]{LopezCorredoira:2002,Reyle:2009,Amores:2017,Uppal:2024}, including classical Cepheids \citep{Skowron:2019a, Skowron:2019b, Chen:2019, Lemasle:2022, Dehnen:2023, CabreraGadea:2024}. For our specific sample of Cepheids, the presence of the warp has been illustrated in Fig. 19 (left panel) of \cite{Skowron:2024}.

In the following Sections, we will infer the large-scale three-dimensional shape of the warp using our two samples of young giants and Cepheids  (Section \ref{sec:warp_fit}), and examine the residuals of the inference (\ref{sec:residuals}). As will be shown below, a classical description of the warp will not be sufficient to explain the observed vertical distribution of stars.

\section{ Analysis of the vertical distribution} \label{sec:analysis}

\subsection{Inference of the warp shape}\label{sec:warp_fit}

\begin{figure*}
\centering
    \includegraphics[width=0.99\textwidth]{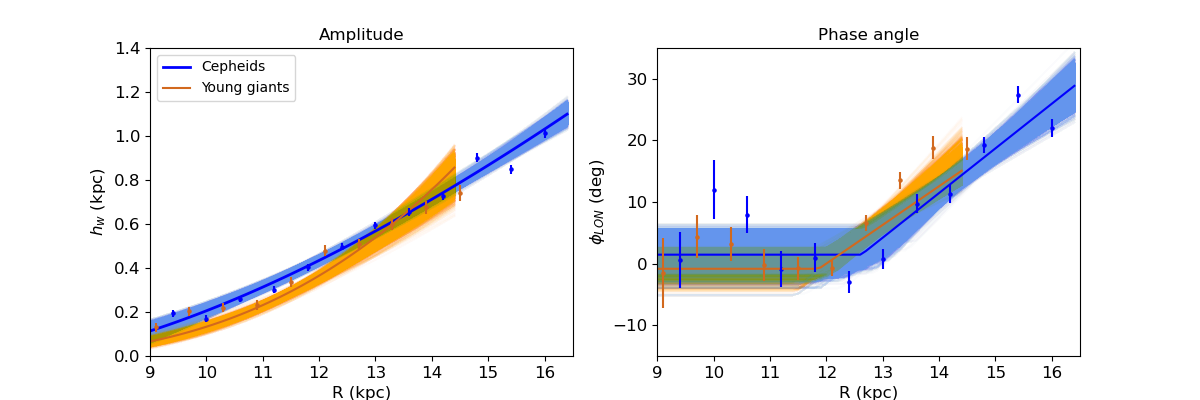}
\caption{ Inferred amplitude (left panel) and phase angle (right panel) for the $m=1$ classical warp model using the young giants sample (orange/brown markers) and the Cepheids sample (blue/light blue markers) based on different approaches. Individual points show the results of the m=1 tilted ring model, while solid lines correspond to the parametric fit, performed on a star-by-star basis, of the twisted line-of-nodes for the young giant (in brown) and Cepheid (in blue) samples. Orange (light blue) transparent lines represent $5\,000$ random realisations of the MCMC chain for the fits of the young giant (Cepheid) sets.   \label{fig:fit_warp_amp_phase_young_giants}
}
\end{figure*}

In this Section, we infer the shape of the Galactic warp using a classical $m=1$ description \citep[similar to][and numerous other works in literature]{Skowron:2019a,Chen:2019}. 
For a given dataset, we will perform the inference using different approaches, to test the robustness of our results. The same procedures will be applied to our two samples of young giants and Cepheids, and the results will be compared to each other. 

We start with a classical 'tilted rings' approach, which we first apply to our young giants sample. We split our dataset into non-overlapping Galactocentric annulii of 0.5 kpc in width. Each annulus typically contains between 900 and 1200 stars. As a test, annulii of different width (ranging from 0.1 kpc to 1.5 kpc) were also considered, to check that the final results of this paper do not depend on the chosen bin size. Stars inside $R=R_{\odot}$ are not considered here, given that the amplitude of the warp has been found to be very small or even null inside that radius by several works in the literature. Young giant stars with $R > 14.5$ kpc are ignored, due to the low number statistics in those regions. For the $i$-th ring, we fit a simple $m=1$ distortion
\begin{equation}
    Z_{w,i} = h_{w,i} \, \sin(\phi - \phi_{LON,i} ) ,
    \label{eq:Zring}
\end{equation}
where $h_{w,i}$ and $\phi_{LON,i}$ are, respectively, the amplitude and phase angle of the line-of-nodes at the radial position covered by the $i$-th ring. For each star, uncertainties in the $Z$-coordinate are computed by converting the entire distribution in heliocentric distance (using 1000 mock resamples for each star), and taken as the 16$^{th}$ and 84$^{th}$ percentiles of the posterior distribution described in Section \ref{subsec:distances}, to a distribution in $Z$, assuming no errors in Galactic latitude $b$. For each ring, we perform the inference using the emcee code \citep{Foreman-Mackey2013}, using a broad uninformative prior for $h_{w,i}$, and imposing that the phase angle $\phi_{LON,i}$ must be within the range $[-180^{\circ},180^{\circ}]$. The likelihood function used to perform the inference is the product of individual Gaussian likelihoods (i.e. sum of log-likelihoods) obtained for each star, assuming a random Gaussian distribution around the model prediction $Z_{w,i}$ from Equation \ref{eq:Zring}, that is
\begin{equation}
\mathcal{L}= \frac{1}{\sqrt{2 \pi} \sigma_z} \, \exp{[-\Delta_Z^2/(2 \sigma_z^2)]} \quad ,
\end{equation}
where $\Delta_Z = Z_{w,i} - Z_i$, and $Z_i$ is the individual vertical coordinate of each star. To describe the total dispersion in the $Z$-coordinate $\sigma_z$, individual uncertainties in $Z$ are summed in quadrature to the intrinsic thickness of the (young) stellar disc, using 150 pc as a reference value (broadly consistent  with the vertical distribution of the sample), although other values have also been tested (see below). The 16$^{th}$ and 84$^{th}$ percentiles of the obtained marginalized posterior distributions on $h_{w,i}$ and $\phi_{LON,i}$ for each ring are shown as brown error bars in Figure \ref{fig:fit_warp_amp_phase_young_giants}, while the best-fit results for each ring are shown as brown points, for the warp amplitude (left panel) and phase angle (right panel) as a function of Galactocentric radius $R$. As pointed out by several works \citep[e.g.][]{Dehnen:2023,CabreraGadea:2024}, the tilted ring approach has the advantage that it does not assume a parametric functional form of the warp parameters (amplitude, shape of the line-of-nodes) as a function for the Galactocentric radius.

As we can see, the warp amplitude systematically increases with $R$. This is expected from previous observations of the large-scale warp. The phase angle, however, varies abruptly from one ring to another inside approximately $R=11 \, \rm{kpc}$, showing large uncertainties. Such behaviour is somehow expected from geometrical considerations: the warp amplitude is very small at these radii \citep[see discussion in][]{Dehnen:2023}, making the measurement of the line-of-nodes more challenging. We notice, however, that there is a net increase of the phase angle in the direction of Galactic rotation  for $R \gtrsim$ 12 kpc, in agreement with \citet{Dehnen:2023} and \citet{CabreraGadea:2024}.

A complementary approach is to infer the warp shape using the individual sources without binning in $R$, but now assuming a parametric shape for the amplitude and phase angle as a function of Galactocentric radius. 
We will refer to this approach as the Global parametric fit. For the amplitude, we assume the typical functional form
\begin{equation}
    h_w(R) =  h_{w,0} (R-R_w)^{\alpha_w}   
\end{equation}
for $R>R_w$, and zero otherwise. We first consider a scenario with a curved (twisted) line-of-nodes (LON), using a functional form that follows the Briggs rule for warps in external galaxies:
\begin{equation*}
\phi_{LON}(R) = \begin{cases}
\phi_{LON,0} &\text{for $R <= R_T$}\\
\phi_{LON,0} + \beta_w \, (R-R_T) &\text{for $R > R_T$} \quad ,
\end{cases}
\end{equation*}
according to which the phase angle is constant in the inner parts of the disc, and curves outside a certain radius $R_T$ (which is by definition greater than the starting radius of the warp $R_W$). The parameter $\beta_w$ defines the rate at which the line-of-nodes changes as a function of $R$ (for $R > R_T$). The final modelled distortion of the Galactic disc along the Z-coordinate is
\begin{equation}
    Z_w(R, \phi) = h_w(R) \, \sin(\phi -\phi_{LON}(R)) \cdot,
    \label{eq:phase_angle}
\end{equation}
which includes the amplitude $ h_w(R)$ and the phase angle $\phi_{LON}$ described above.
The results of the statistical inference are presented in Table \ref{table:fit_young_giants}.
Systematic uncertainties on each parameter were estimated by testing five different values for the thickness of the disc, based on the typical thickness of young stellar populations, namely 100, 150, 200, 250, and 300 pc. The range of the derived warp parameters assuming different values of disc thickness is adopted as the systematic uncertainty on each specific parameter.

%
\begin{table}
\caption{Best-fit results for the m=1 global parametric warp model with a twisted line-of-nodes (see text) using the young giants sample.}             
\label{table:fit_young_giants}     
\centering                          
\renewcommand{\arraystretch}{1.3}
\begin{tabular}{c c c c}       
\hline\hline  
Parameter & Best-fit & Statistical & Systematic  \\   
          &  value & uncertainty & uncertainty  \\  
\hline\hline                       
$R_w$ (kpc) & 5.5 & $^{+0.8}_{-0.7}$ & $^{+1.3}_{-0.8}$ \\ 
$h_{w,0}$ (kpc$^{ \alpha_w-1}$) & 0.012 & $^{+0.002}_{-0.001}$ & $^{+0.006}_{-0.001}$ \\ 
$\alpha_w$ & 1.9 & $^{+0.3}_{-0.3}$ & $^{+0.5}_{-0.6}$ \\ 
$\phi_{LON,0}$ (deg) & 0.1 & $^{+0.6}_{-0.6}$ & $^{+1.1}_{-1.2}$\\ 
$R_t$ (kpc) & 12.1 & $^{+0.1}_{-0.1}$ & $^{+0.1}_{-0.1}$\\ 
$\beta_w$  (deg/kpc) & 9.9 & $^{+0.5}_{-0.4}$ & $^{+0.4}_{-0.2}$\\ 
\hline                                  
\end{tabular}
\end{table}

%
%
\begin{table}
\caption{Best-fit results for the m=1 global parametric warp model with a twisted line-of-nodes (see text) using the Cepheids sample.}             
\label{table:fit_Cepheids}      
\centering                         
\renewcommand{\arraystretch}{1.3}
\begin{tabular}{c c c c}        
\hline\hline  
Parameter & Best-fit & Statistical & Systematic  \\    
          &  value & uncertainty & uncertainty  \\  
\hline\hline                       
$R_w$ (kpc) & 7.7 & $^{+0.3}_{-0.4}$ & $^{+0.1}_{-0.1}$ \\ 
$h_{w,0}$ (kpc$^{ \alpha_w-1}$) & 0.057 & $^{+0.012}_{-0.011}$ & $^{+0.005 }_{-0.001 }$ \\ 
$\alpha_w$ & 1.3 & $^{+0.1}_{-0.1}$ & $^{+0.1 }_{-0.1}$ \\ 
$\phi_{LON,0}$ (deg) & 0.9 & $^{+1.1}_{-1.1}$ & $^{+0.4}_{-0.2}$\\ 
$R_t$ (kpc) & 12.6 & $^{+0.2}_{-0.2}$ & $^{+0.1}_{-0.1}$\\ 
$\beta_w$  (deg/kpc) & 8.0 & $^{+0.5}_{-0.5}$ & $^{+0.2}_{-0.2}$\\ 
\hline                                  
\end{tabular}
\end{table}
%

We note that, based on this approach, the starting radius of the warp $R_W$ was treated as a free parameter, and we obtained a very low value of $R_w$. We should, however, avoid over-interpreting such a result: it is clearly evident from Figure \ref{fig:fit_warp_amp_phase_young_giants} (left panel) that the warp amplitude at $R$=9 kpc is smaller than the typical vertical scale-height of the sample. Different mathematical prescriptions with different values of $R_w$ can lead to very similar warp amplitudes in the inner disc (e.g. $R \lesssim$ 8-10 kpc), whose differences are presumably too small to be detected, taking into account the intrinsic thickness of the Galactic disc. We note that, for the young giants sample, the tilted rings method and the parametric global fit give very similar results, as shown in Figure \ref{fig:fit_warp_amp_phase_young_giants}.

We note that if instead a straight LON (i.e. with a constant phase angle) is assumed, so that Equation \ref{eq:phase_angle} becomes $\phi_{LON}= \phi_{LON,0}$ at all radii, the resulting warp amplitude of the young giants is similar to the twisted LON scenario, but now with a phase angle $\phi_{LON}=9.9{\degree} \pm 0.6{\degree}$. 

The analysis performed with the young giants catalog was also repeated with our Cepheids sample. Figure \ref{fig:fit_warp_amp_phase_young_giants} shows the results obtained with the m=1 non-overlapping tilted rings method (blue circles). With this sample, due to the relatively low-number statistics, we increased the width of the rings to 0.6 kpc in Galactocentric radius. Each ring typically contains between 100 and 230 stars. Figure \ref{fig:fit_warp_amp_phase_young_giants} shows the results for $R>9$ kpc, where the warp signal is more apparent. We do not consider stars outside $R \gtrsim 16.5$ kpc and/or on the other side of the Galaxy, because the number of objects is too small to give reliable results. We note that this sample reaches much larger distances than the young giant sample. For this sample, we also apply the parametric global fit with a twisted LON, obtaining the results presented in Table \ref{table:fit_Cepheids}. For the $R_w$ parameter, the same considerations made for the young giant sample are also valid for the Cepheid sample. We also test a straight LON model, obtaining a phase angle $\phi_{LON}=14.0{\degree} \pm 0.6{\degree}$.

The results obtained for the warp structure with our two samples are in good agreement with each other. This is somehow expected, because they are both tracing the young stellar populations in the Galactic disc. 
This is also in very good agreement with previous works based on young stars. For instance, \cite{Dehnen:2023} find an inclination angle of 3$^{\circ}$ at $R \geq$ 14 kpc, which corresponds to an amplitude of 0.7 kpc at $R=14$ kpc, in perfect agreement with the left panel of Figure \ref{fig:fit_warp_amp_phase_young_giants}. Moreover, we note that the general behaviour of the line-of-nodes in our two samples is in agreement with the results obtained in the literature with classical Cepheids \citep[][]{Dehnen:2023, CabreraGadea:2024}.

\subsection{Analysis of the residuals}\label{sec:residuals}

\begin{figure*}
\centering
    \includegraphics[width=1\textwidth]{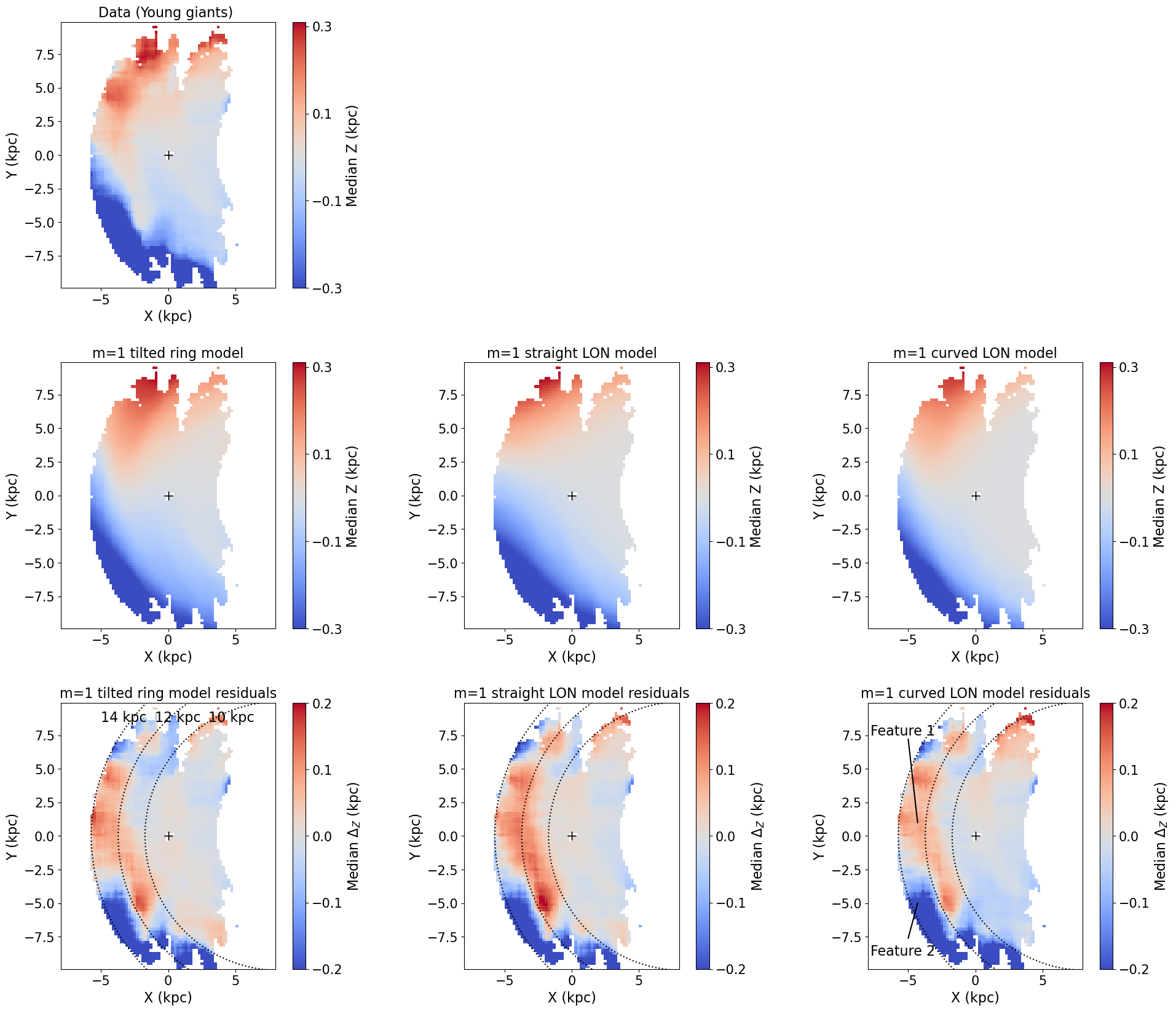}
\caption{ Comparison between the maps of median $Z$-coordinate from observations, best-fit model and residuals. \emph{ Upper Left panel}: median Z-coordinate of the young giant sample. The position of the Sun is shown by the black cross at (X,Y)=(0 kpc, 0 kpc). 
\emph{Middle panels}: prediction from the m=1 tilted rings model (left), straight line-of-nodes parametric model (center) and twisted line-of-nodes parametric model (right) (right). \emph{Lower panels}: corresponding vertical residuals for the m=1 tilted rings model (left), straight line-of-nodes parametric model (center) and twisted line-of-nodes parametric model (right). Dotted lines in the three lower panels show rings at constant radii $R$=10, 12 and 14 kpc from the Galactic center. 
\label{fig:data_model_residuals_young_giants}
}
\end{figure*}

\begin{figure*}
\centering
    \includegraphics[width=1\textwidth]{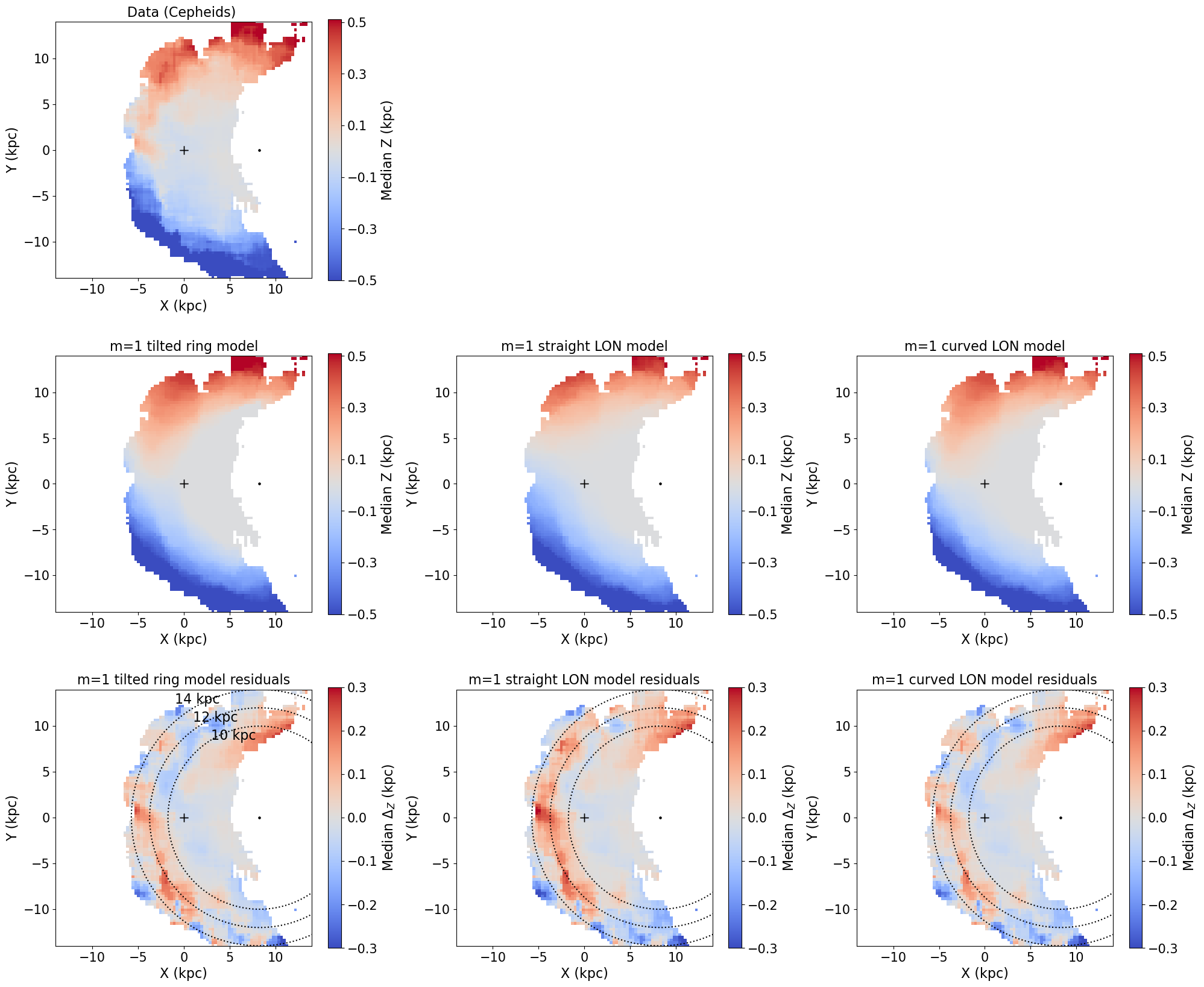}
\caption{ Same as Figure \ref{fig:data_model_residuals_young_giants}, but now using the Cepheids sample. The black cross shows the position of the Sun, while the black dot shows the position of the Galactic center. Due to the different spatial coverage, the range in Z is different compared to Figure \ref{fig:data_model_residuals_young_giants}. \label{fig:data_model_residuals_Cepheids}
}
\end{figure*}

\begin{figure*}
\centering
    \includegraphics[width=1\textwidth]{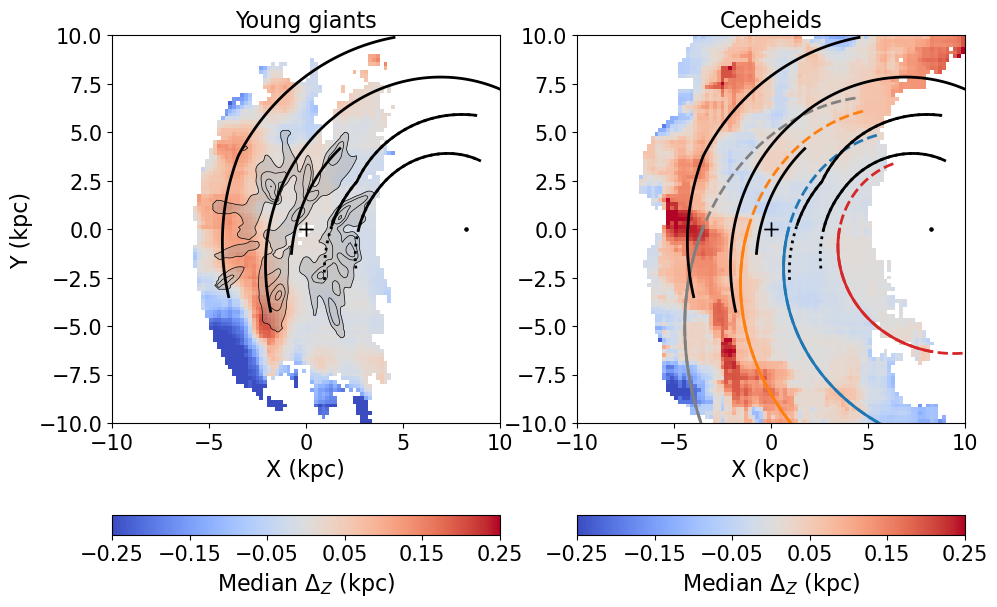}
\caption{ Vertical residuals $\Delta_Z$ of the m=1 straight line-of-nodes model using young giants (left panel) and Cepheids (right panel), compared to spiral arm geometries taken from literature: \cite{Reid:2019} (black solid lines, both panels), \cite{Poggio:2021} (grey shaded contours, left panel) and \cite{Drimmel:2024} (colored lines, right panel). For the \cite{Drimmel:2024} model, different colors indicate the Perseus arm (grey), Local/Orion arm (orange), Sagittarius-Carina arm (blue) and Scutum arm (red). 
\label{fig:comparison_zresiduals_spiral_arms}
}
\end{figure*}

\begin{figure*}
\centering
    \includegraphics[width=1\textwidth]{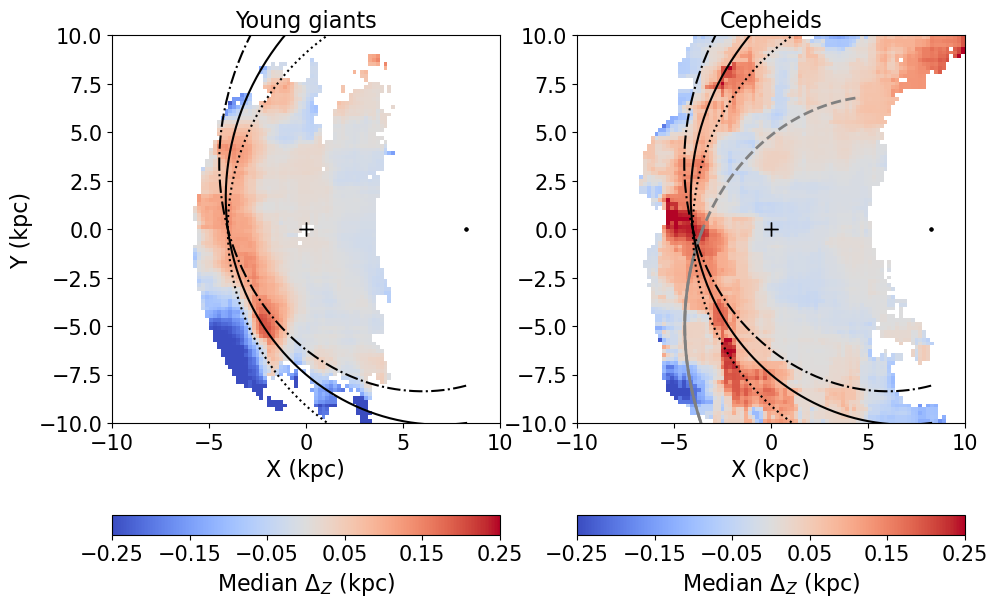}
\caption{ Same as Figure \ref{fig:comparison_zresiduals_spiral_arms}, but now compared to different curves, plotted for visual reference. The dashed-dotted and solid black lines show a leading spiral curve with pitch angles -15$\degree$ and -7.5$\degree$, respectively (see text for more details). The black dotted line shows a circle (pitch angle 0$\degree$) at constant radius $R$=12.3 kpc. In the right panel, the Perseus arm from \cite{Drimmel:2024} (grey line) is also shown for reference.
\label{fig:comparison_zresiduals_leading_spiral_curve}
}
\end{figure*}

It is natural to wonder whether the classical $m=1$ description adopted in the previous Section is sufficient to explain the three-dimensional distribution observed in our datasets. Several works in the literature have already pointed out that a simple $m=1$ model is not enough to explain the warp in our Galaxy, suggesting that it is lopsided \citep{Romero:2019}, or even fitting a Fourier decomposition with $m=0,1,2$ terms \citep{Levine:2006, CabreraGadea:2024}.
The approach presented here is complementary to previous works: we fit a simple $m=1$ term, and then investigate the residuals, with the aim of determining which additional ingredient is needed to explain the observed vertical structure of the young Galactic disc.

Figure \ref{fig:data_model_residuals_young_giants} shows the comparison between the median Z in the Galactic plane observed in the young giant dataset (upper left panel), the prediction from the tilted ring model (left panel, middle row), straight line-of-nodes (LON) model (middle panel, middle row) and the twisted LON model (right panel, middle row) described in Section \ref{sec:analysis} (middle panel), and the corresponding residuals (lower panels). As we can see, the residuals of the fit show a clear coherent pattern rather than random noise. The pattern is similar for all three m=1 models considered here. Specifically, we can identify a specific region of the outer Galactic disc where the distribution of stars is systematically shifted toward positive Z compared to the global best-fit model (i.e. red region labelled as Feature 1), extending from approximately $(X,Y) \simeq$ ($-4$ kpc, 4 kpc) to $(X,Y) \simeq$ ($-2$ kpc, $-4$ kpc). The morphology of the observed feature is consistent with a nearly radial ripple or corrugation, which cannot be explained by a simple $m=1$ warp. The observed corrugation reaches about 150-200 pc in vertical height, and extends over the entire volume covered by the young giant sample for about 10 kpc (corresponding to about 50 degrees in Galactic azimuth). On the other hand, no major vertical displacement is detected inside Galactocentric radii of 8-9 kpc \citep[see also][]{Xu:2015,Hunt:2022}, though it is possible that another ripple with very small vertical amplitude is present close to the Solar circle (we defer the study of its possible existence to future studies).

In the lower left corner of the residual maps (lower panels), we can see that stars are systematically located lower than the prediction from our models (blue region labelled as Feature 2). It is impossible not to wonder whether the observed blue region is the 'negative counterpart' of the red region observed at inner Galactic radii, being both part of the same oscillatory pattern. We note, however, that the blue region is at the border of our map, and therefore we treat it with caution. 

To test the robustness of the obtained maps, we re-performed the warp inference using only the subset of stars with high quality distances (i.e. based on a good parallax signal-to-noise ratio, $\varpi / \sigma_{\varpi} > 5$), or using other distance estimates than those calculated in Section \ref{subsec:distances} \citep[for instance from][or selecting stars with $\varpi / \sigma_{\varpi} > 5$ and calculating the distance as 1/parallax]{BailerJones:2021,Anders:2022}, always obtaining residuals maps with features similar to those shown in Figure \ref{fig:data_model_residuals_young_giants}.  

Figure \ref{fig:data_model_residuals_Cepheids} shows again a comparison between the data (upper left panel), the best-fit predictions from our models (middle panels) and the corresponding residuals (lower panels), similar to Figure \ref{fig:data_model_residuals_young_giants}, but now for our sample of Cepheids. We note that the lower panels in Figure \ref{fig:data_model_residuals_Cepheids} exhibit some similarities and differences (and there are several reasons for this being so, as discussed in the following). The residuals of the tilted ring model (lower left panel) and the twisted LON model (lower right panel) exhibit a region of systematically positive residuals (red area) in the lower parts of the plot (Y $\lessapprox$ 4 kpc), for Galactocentric radii between 10 and 14 kpc. In the residuals of the straight LON model (lower central panel), the observed feature is even more prominent, with systematically positive residuals that extend from $Y \sim$ 10 kpc to $Y \sim$ - 10 kpc in the range of Galactocentric radii 10-14 kpc. For all three considered models, the red region detected in the residuals of the Cepheids sample might correspond to the corrugation (labelled as `Feature 1') found with the young giant sample, suggesting a similar behaviour. 

To enable a proper comparison between our two samples, the observed residuals are plotted on the same XY-range in Figure \ref{fig:comparison_zresiduals_spiral_arms}. As we can see, the red region of systematically positive residuals discussed above is located at similar positions in the Galactic plane for our two samples. To better explore the geometry of the observed residuals, our maps are then compared to spiral arms models available in literature. For the Cepheids sample, we overplot the model of \cite{Drimmel:2024} (colored lines in the right panel of Figure \ref{fig:comparison_zresiduals_spiral_arms}), which is derived from the same dataset used here. For both the young giant and Cepheids sample, we overplot the spiral arm model from \cite{Reid:2019} (black solid lines), which is based on masers. Finally, for the young giant sample, we also overplot the overdensity contours of the upper main sequence stars from \cite{Poggio:2021} (grey shaded areas in the left panel of Figure \ref{fig:comparison_zresiduals_spiral_arms}), which we found to be in good agreement with the spiral structure observed in the young giant sample (Figure \ref{fig:xy_map_young_giants}). The general picture emerging from Figure \ref{fig:comparison_zresiduals_spiral_arms} is that the vertical residuals observed in this work do not exhibit an obvious correlation with the different spiral arm geometries considered here: unlike the red region in the vertical residuals, the spiral arms tend to have a trailing geometry (i.e. positive pitch angle), so that they are opening in the direction opposite to Galactic rotation. We note, however, that the outer arm from the model of \cite{Reid:2019} (outermost black solid line in Figure \ref{fig:comparison_zresiduals_spiral_arms}) has a very small pitch angle (specifically, 3$\degree$ in the region of overlap with our maps), which implies that that arm is almost a radial feature. While there is a partial overlap in some regions (approximately $-2.5$ kpc $<Y< 5$ kpc), the orientation of the red feature in the vertical residuals is different, because it is more similar to a leading (i.e. negative pitch angle) rather than a trailing (i.e. positive pitch angle) spiral.  We also compared our maps to the spiral arm model from \cite{Hou:2014} (here not shown for brevity), reaching conclusions similar to those obtained with the spiral models discussed above. 

It is already evident from the lower panels of Figure \ref{fig:data_model_residuals_young_giants} and \ref{fig:data_model_residuals_Cepheids} that the corrugation is not exactly radial: in the lower parts of the plot ($Y \lessapprox -2$ kpc) it tends to be within 10 and 12 kpc in Galactocentric radius $R$, while it is located between 12 and 14 kpc in the upper parts ($Y \gtrapprox -2$ kpc). For visual reference, Figure \ref{fig:comparison_zresiduals_leading_spiral_curve} shows a comparison between our maps and two leading spiral curves with pitch angles of -15$\degree$ (dashed-dotted black curve) and -7.5$\degree$ (solid black curve), which cross the radius $R$=12.3 kpc at Galactic azimuth $\phi=0 \degree$ (toward the Galactic anticenter). The black dotted line shows a curve at constant radius $R$=12.3 kpc, shown for reference.

\section{Kinematics}\label{sec:kinematics}

\begin{figure*}
\centering
\includegraphics[width=1\textwidth]{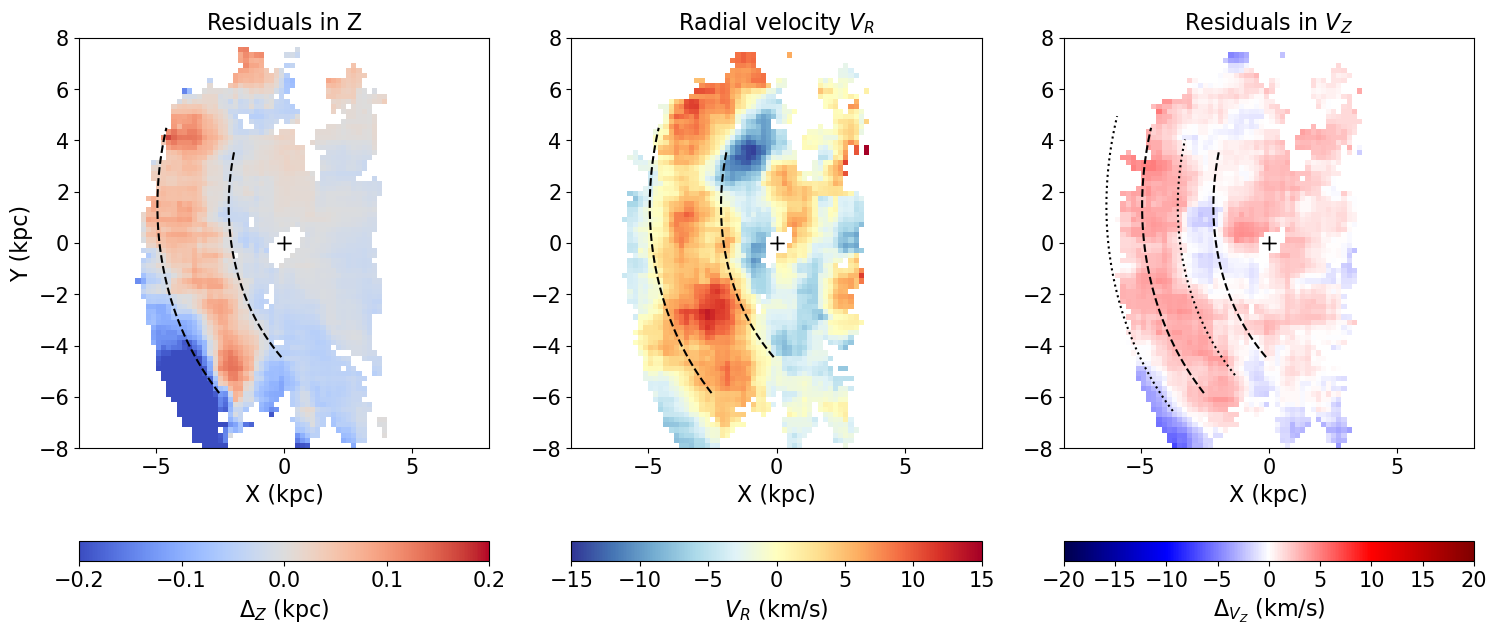}
\caption{ Comparison between the residuals in the vertical coordinate $\Delta_Z = Z - Z_{WARP}$ (left panel), the median radial motions $V_R$ (middle panel) and the residuals in the vertical velocity $\Delta_{VZ} = V_Z - V_{Z,WARP}$ for the young giant sample. Dashed lines give a rough indication of the region where positive residuals $\Delta_Z$ are obtained with this sample. Dotted lines are the same, but shifted by half the amplitude of the region (see text). \label{fig:scatter_Zres_VR_Vzres_young_giants}
}
\end{figure*}

\begin{figure*}
\centering
\includegraphics[width=1\textwidth]{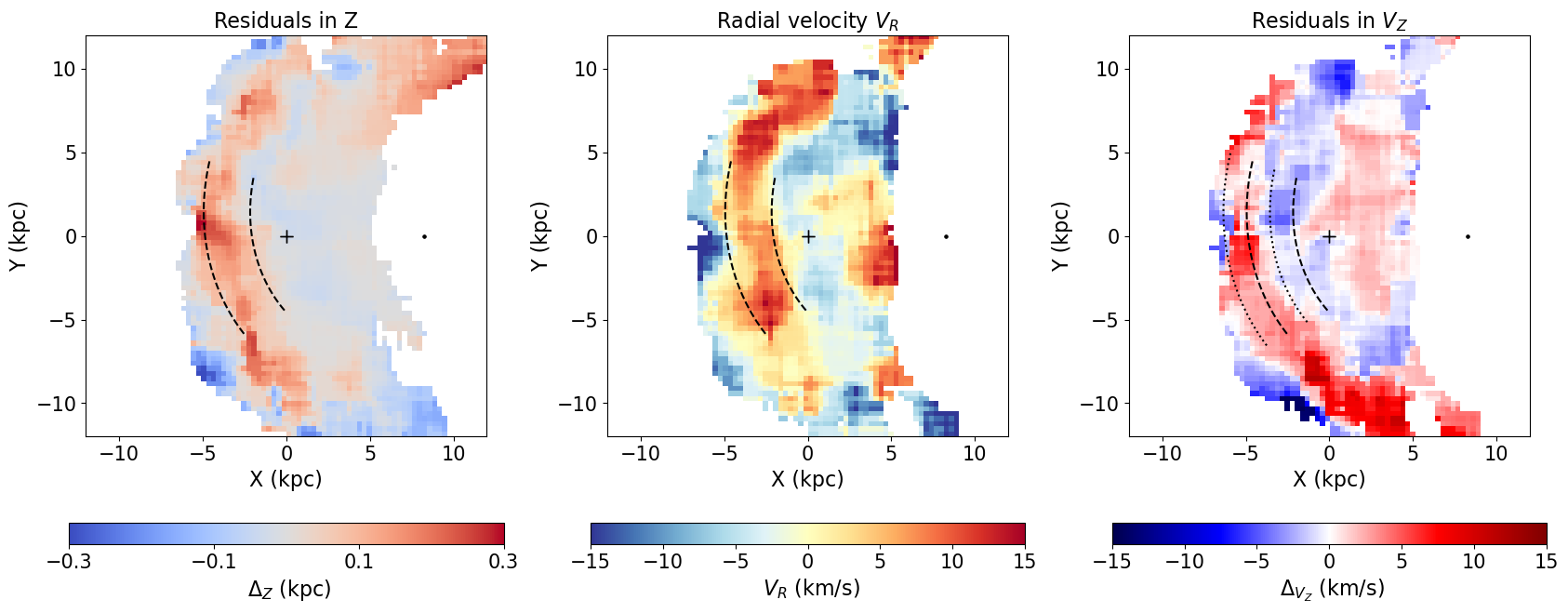}
\caption{Same as Figure \ref{fig:scatter_Zres_VR_Vzres_young_giants}, but now for the Cepheids sample. \label{fig:scatter_Zres_VR_Vzres_Cepheids}
}
\end{figure*}

\begin{figure*}
\centering
\includegraphics[width=0.9\textwidth]{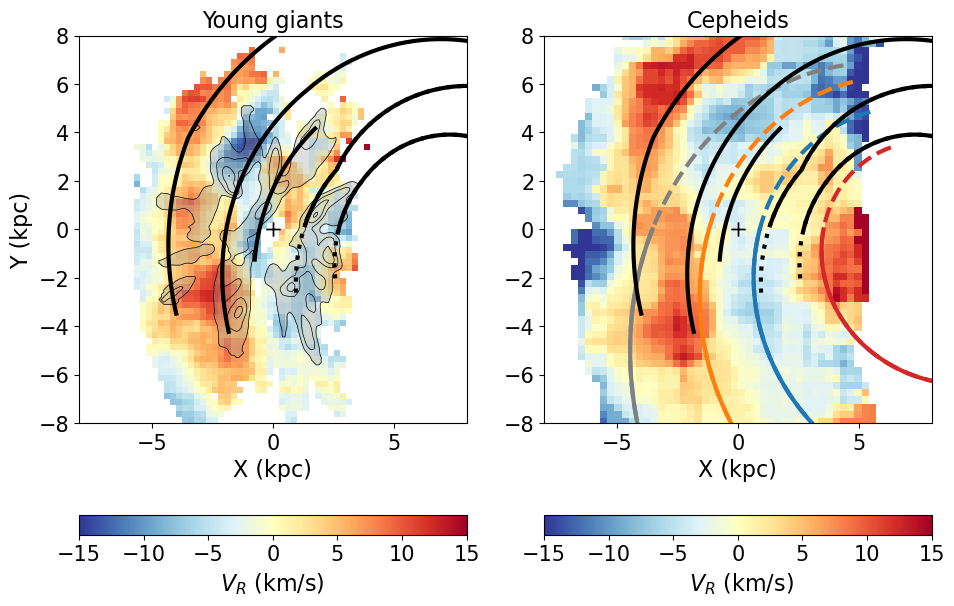}
\caption{ Face-on view of the median radial motions $V_R$ for the young giant sample (left panel) and the Cepheids sample (right panel), compared to different spiral arm geometries from literature: \cite{Reid:2019} (black solid lines, both panels), \cite{Poggio:2021} (grey shaded contours, left panel) and \cite{Drimmel:2024} (colored lines, right panel). For the \cite{Drimmel:2024} model, different colors indicate the Perseus arm (grey), Local/Orion arm (orange), Sagittarius-Carina arm (blue) and Scutum arm (red). \label{fig:VR_spiral_arms} 
}
\end{figure*}

\begin{figure}
\centering
    \includegraphics[width=0.50\textwidth]{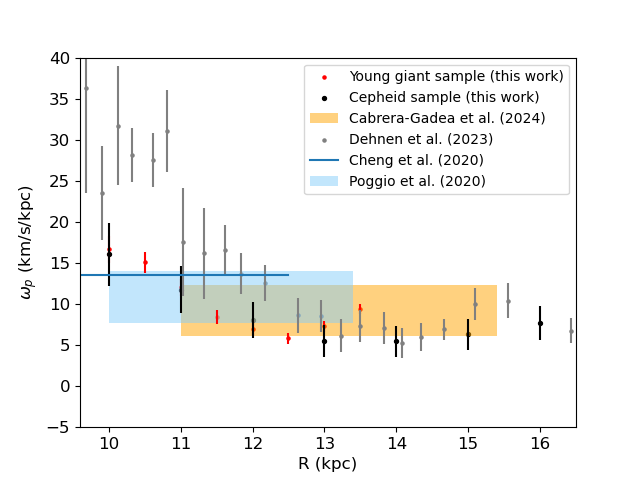}
\caption{ Comparison between the obtained values of warp precession with our young giants (red points) and Cepheids sample (black points), together with other values from the literature. \label{fig:warp_precession}
}
\end{figure}

\subsection{Radial motions}

To learn more about the residuals observed in the vertical distribution of stars, we now compare them to the stellar velocity field, looking for possible correlations. In the left panel of Figure \ref{fig:scatter_Zres_VR_Vzres_young_giants}, we can see the same residuals shown in Figure \ref{fig:data_model_residuals_young_giants} (right panel) for the young giant sample, now overplotted with two dashed lines that roughly delimit the region where the majority of the positive residuals in Z are located (i.e. Feature 1 discussed above), indicated simply for the purpose of comparison with the velocity maps of the young giants (as well as the Cepheid sample).

The middle panel of Figure \ref{fig:scatter_Zres_VR_Vzres_young_giants} shows the median radial component of the velocity $V_R$ in Galactocentric cylindrical coordinates for the young giant sample. We note that the $V_R$ map shown here does not depend on any warp model. This is because the warp is mainly a vertical distortion. In our coordinate system, positive (negative) values of $V_R$ correspond to velocities directed towards the outer (inner) parts of the Galactic disc. Surprisingly (at least to us), the stars located in Feature 1 (i.e. between the dashed lines) are typically moving outwards, with  velocities of about 10-15 km/s. Roughly outside the dashed lines, the typical orientation of the velocities changes in sign, with stars that are typically moving inwards.

A similar pattern is also observed in the median $V_R$ of our Cepheids sample, shown in the middle panel of Figure \ref{fig:scatter_Zres_VR_Vzres_Cepheids}. As we can see, also with this sample the positive $V_R$ region coincides with the positive crest identified with the young giants (delimited by the dashed lines). Furthermore, the Cepheids sample indicates that such features {\bf is} even more extended, reaching about 20 kpc in length.

To better quantify the statistical correlation between the $\Delta_Z$ and $V_R$ maps, we dissected the XY-plane in large (non-overlapping) pixels of 1 kpc (2 kpc) width for the young giants (Cepheids) sample. The median $\Delta_Z$ and $V_R$ have been calculated for the pixels that contain at least 10 stars from the young giant sample (5 stars from the Cepheids catalog), and only those with R>10.5 kpc (where the above-discussed features are located). To quantify the correlation between these two quantities, we calculated Kendall's correlation coefficient \citep{Kendall:1938}, obtaining values between $\tau_b=0.33$ and 0.37 for the young giants sample, and between 0.3 and 0.46 for the Cepheids sample, depending on which of the three different $m=1$ warp models was considered (with the straight LON model showing larger correlation). In all three cases, and for both samples, the obtained $\tau_b$ indicates a strong positive correlation\footnote{A correlation is typically classified as very weak for values less than 0.10, weak between 0.10 and 0.19, moderate between 0.20 and 0.29, and strong for 0.30 or above \citep{botsch2011chapter}.}, suggesting that regions of the disc with systematically positive residuals in Z (i.e. shifted upwards with respect to a simple $m=1$ warp model) statistically tend to exhibit positive values of $V_R$, and vice versa.

The observed statistical correlation might suggest that the quantities $\Delta_Z$ and $V_R$ are linked by a common physical origin (see discussion in Section \ref{sec:discussion}). However, correlation doesn't necessarily imply causation. It is therefore important to consider alternative scenarios, like, for instance, the possible impact of the spiral structure on the in-plane motions. Figure \ref{fig:VR_spiral_arms} (right panel) shows the comparison between the $V_R$ map and the spiral model from \cite{Drimmel:2024}. As we can see, the observed $V_R$ features do not exhibit an obvious aligment with this spiral model. Specifically, the observed large-scale positive $V_R$ feature dominating the outer (>$R_{ /odot}$) disc is crossing the Perseus arm (grey curve) and the Local Arm (orange curve). A similar consideration can be done while comparing the $V_R$ map of the young giant sample with the overdensity contours from \cite{Poggio:2021} (which, from left to right, show segments of the Perseus, the Local and the Sagittarius-Carina arm) in the left panel of Figure \ref{fig:VR_spiral_arms}. Again, a large-scale positive $V_R$ feature in the outer disc is connecting the Perseus and the Local Arm. The overdensity in the upper part of the Perseus arm (the Cassiopeia region, at approximately (X,Y)$\sim$(-2 kpc, 3 kpc)) systematically moves inwards (negative $V_R$), showing a different behaviour compared to the rest of the Perseus arm. Similarly, the Local Arm exhibit alternate regions of inwards/outwards motions. On the other hand, the Sagittarius-Carina arm is mostly dominated by inward motions, especially in the lower parts of the plot). The spiral arms from \cite{Reid:2019} (solid black lines) have a different orientation with respect to the other previously considered spiral arms geometries. In the outer disc, the positive $V_R$ feature does not exhibit an obvious alignment with the Perseus arm from this model, while there is an overlap with the outer arm (left black curve), though with a sligtly different orientation: in the upper part of the plot, the outer arm seems to be located in the portion of the red $V_R$ feature which is closer to the Galactic center (GC), while in the lower parts of the plot the outer arm is at the outer edge (i.e. more distant parts from the GC) of the positive $V_R$ feature. 

\subsection{Vertical motions}

If the residuals discussed in Section \ref{sec:residuals} are related to a vertical wave in the Galactic disc, we should see a trace in the vertical velocities $V_Z$ of the stars of our samples. However, the vertical velocity field of stars in the Galactic disc is expected to be dominated by the signature of the large-scale warp \citep[e.g.][]{Schoenrich:2018, Poggio:2018, Romero:2019}. Therefore, the natural approach would be to remove the signature expected from the Galactic warp, and then investigate the residuals (similar to the approach performed with the Z-coordinate). 

However, modelling the warp signature is not trivial. Several works have found that, in addition to the spatial parameters that describe the three-dimensional shape of the warp, at least one kinematic parameter is needed \citep{Poggio:2020,Cheng:2020, Dehnen:2023, HrannarMcMillan2024, CabreraGadea:2024, Zhou:2024}. Using classical Cepheids \cite{Dehnen:2023} and \cite{CabreraGadea:2024} found a prograde precession that decreases as a function of Galactocentric radius $R$.

Figure \ref{fig:warp_precession} shows the results obtained for the young giants (red points) and Cepheids sample (black points) after dividing our samples in radial bins, and fitting for the warp precession $\omega_p$ separately (adopting the warp best-fit models discussed in Section \ref{sec:analysis}). With the young giant sample, the bins are of 1.5 kpc in width, and typically contain between 2500 and 3000 stars. For the Cepheid sample, due to the low number of stars with available line-of-sight measurements, we chose a relatively large bin size of 4 kpc, allowing us to have approximately 500-800 stars for each bin.
We infer the warp precession using Equation 9 in \citet{Cheng:2020}, which also includes the mean radial motions $V_R$ (although we obtain very similar results also if the $V_R$ term is not included). As we can see, our results are in good agreement with other works from the literature, especially for $R \gtrapprox 11$ kpc, where the warp amplitude is larger (see previous Sections). 

After subtracting the prediction from the above-described kinematic model, the residual vertical velocities $\Delta_{V_Z}$ are shown in the right panels of Figures \ref{fig:scatter_Zres_VR_Vzres_young_giants} and  \ref{fig:scatter_Zres_VR_Vzres_Cepheids}, respectively for the young giants and the classical Cepheids. In both plots, we see a feature with systematically positive $\Delta_{V_Z}$. This feature is similar to the one observed in the $V_R$ maps and $\Delta_{Z}$, but now slightly shifted towards the outer parts of the disc. 
For comparison, we plot the same lines shown in the previous plots, as well as two additional lines, now shifted by 1.4 kpc (which is the half distance between the two dashed lines), shown as dotted lines. 
It is interesting to note that the new dotted lines roughly contain systematically positive $\Delta_{V_Z}$. 

To test the robustness of the obtained results against different warp kinematic models, we tested different values of $\omega_p$=12, 10, 8 and 3 km/s/kpc, and checked the corresponding $\Delta_{V_Z}$ maps. As expected, we found that the $\Delta_{V_Z}$ structure associated with the corrugation was always present, though with a different magnitude in $\Delta_{VZ}$.

\begin{figure*}
\centering
    \includegraphics[width=0.95\textwidth]{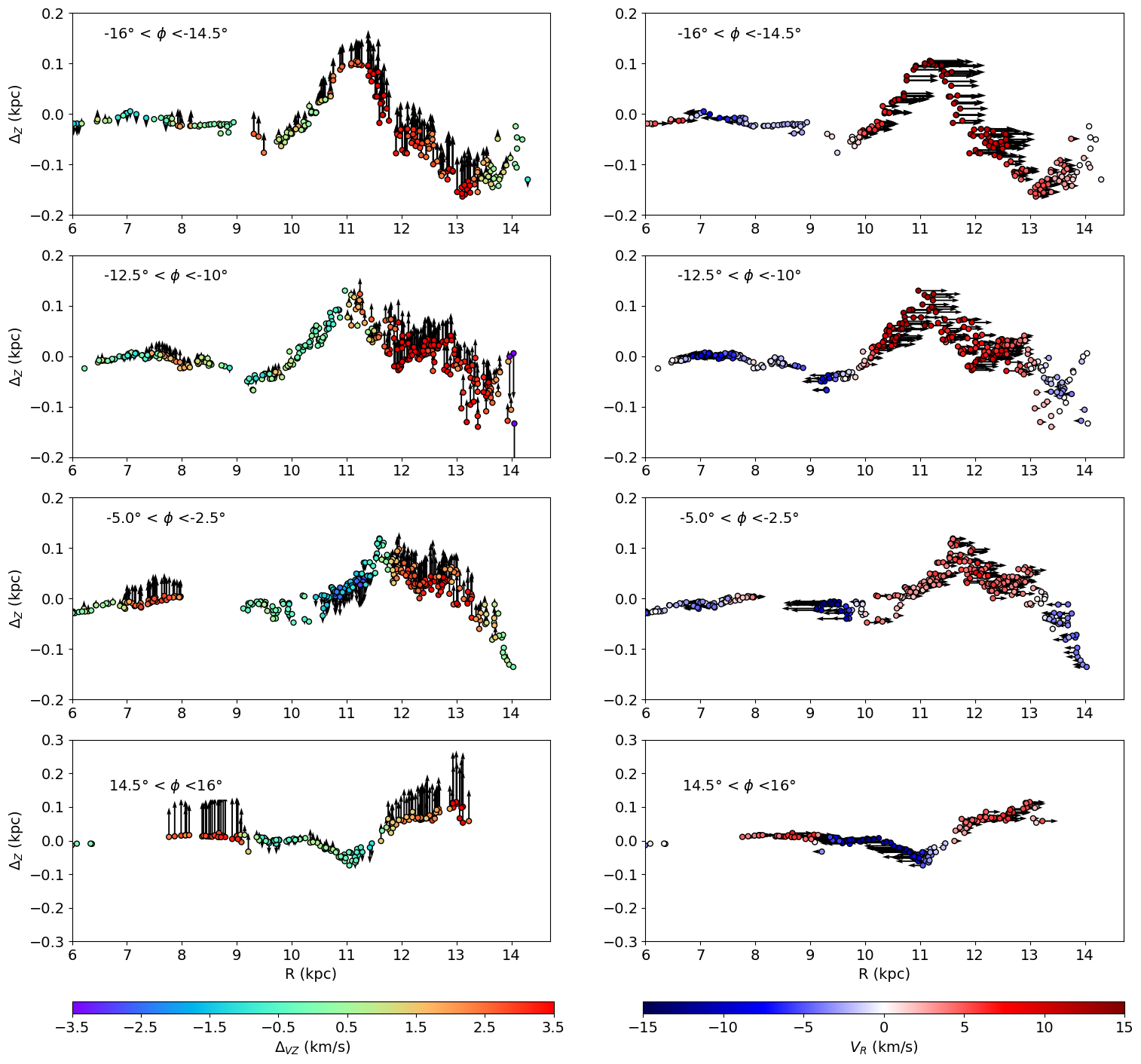}
\caption{ Edge-on view of the detected corrugation using the young giant sample. \emph{Left columns:} vertical spatial residuals $\Delta_Z$ as a function of Galactocentric radius $R$ for different slices in Galactic azimuth $\phi$, as indicated by the individual labels. Points are color-coded by the residuals in the vertical velocity $\Delta_{VZ}$. Black arrows show the direction and magnitude of the median residual vertical velocity $\Delta_{VZ}$. \emph{Right columns:} same as left panels, but now color-coded by the median radial velocity $V_R$. Black arrows show the direction and magnitude of the median radial velocity $V_R$. \label{fig:vector}
}
\end{figure*}
 
\section{Discussion}\label{sec:discussion}

\begin{figure*}
    \centering
    \includegraphics[width=1\textwidth]{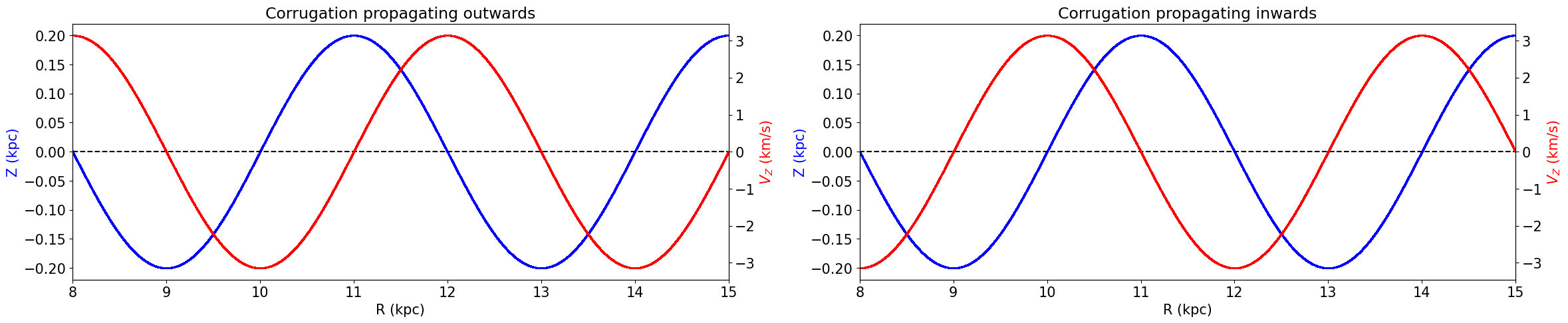}
    \caption{ Predicted $Z$ coordinate (blue) and $V_Z$ pattern (red) as a function of Galactictocentric radius R based on the simple toy model described in Appendix \ref{sec:toy_model} (see Equation \ref{eq:ZC_corrugation}). The curves show the case of a corrugation propagating towards the outer parts of the disc at $v_c$=10 km/s (left panel) or the inner parts of the disc at $v_c$=-10 km/s (right panel).}
    \label{fig:z_vz_1D}
\end{figure*}

\begin{figure}
    \centering
    \includegraphics[width=0.5\textwidth]{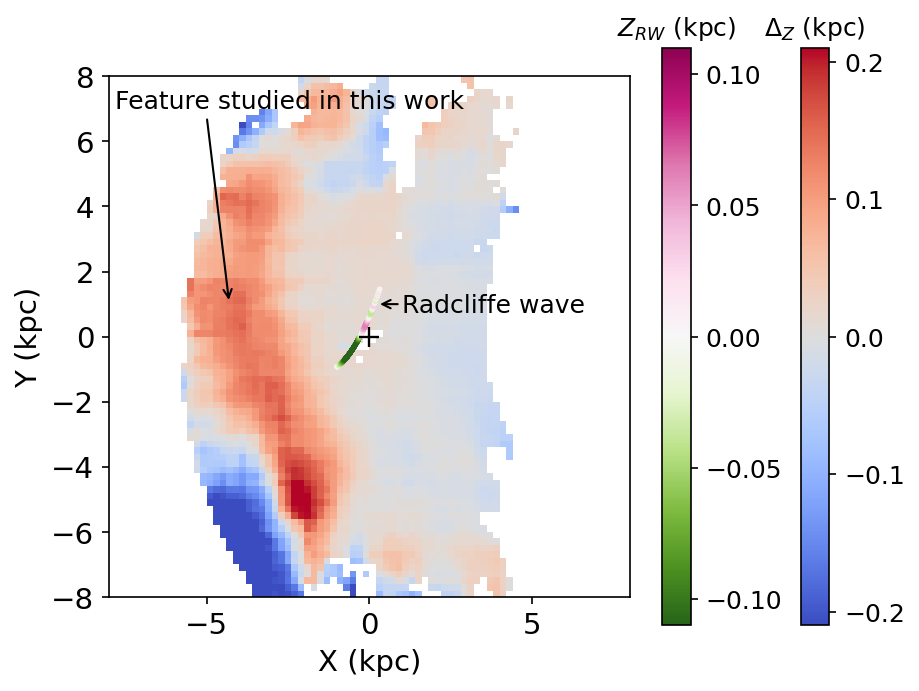}
    \caption{ Comparison between the feature studied in this work and the Radcliffe wave \citep{Alves:2020}, based on the model of \cite{Konietzka:2024}.}
    \label{fig:Radcliffe}
\end{figure}

To facilitate the interpretation of the obtained results, we compared the observed maps to a simple toy model describing a vertical corrugation propagating in the Galactic stellar disc. The toy model, based on the $0^{th}$ moment of the Collisionless Boltzmann Equation (see details in Section \ref{sec:toy_model}), is not designed to provide a sophisticated and realistic description of the Milky Way, but is simply a first step towards understanding the results obtained in this work. Figure \ref{fig:z_vz_1D} (left panel) shows the prediction for a corrugation propagating towards the outer disc, which would manifest itself with a vertical velocity  pattern (red curve) shifted in radius $R$ by about half the radial width of the corrugation towards the outer disc with respect to the vertical displacement in $Z$ (blue curve). The opposite behaviour would be expected for a corrugation moving towards the inner Galaxy (i.e. with $V_Z$ features shifted towards the inner disc with respect to the $Z$ pattern, see the right panel of Figure \ref{fig:z_vz_1D}). It is important to note that the toy model is simply describing the expected kinematic signature for a given type of corrugation propagating in the Galactic disc, and does not explain in any way which phenomenon is at the origin of the observed perturbation. Thus, based on this simple model, a possible interpretation for our observed maps is that a vertical wave extending over a large portion of the Galactic disc
 is present, and it is propagating outwards. The spatial manifestation of such a wave would be a radial corrugation on top of the large-scale Galactic warp, which we detect in the vertical residuals $\Delta_Z$. Future work based on more realistic models will help us improve our understanding of the propagation of waves in the Milky Way disc.

The pattern found in our datasets is in good agreement with the numerical simulations analysed in \cite{Gomez:2013}, which exhibit vertical perturbations of the Milky Way disc excited by a Sagittarius-like satellite galaxy. In their Figure 6 (Panel E), they find that where the mean $<Z>$ of the Galactic disc takes an extrema, $<V_Z> \sim 0$ (similar to the toy model in Figure \ref{fig:z_vz_1D}). They interpret this oscillatory behaviour as the signature of a vertical wave.

Using N-body simulations of a Milky Way-like galaxy perturbed by a dwarf galaxy, \cite{Price-Whelan:2015} showed that, in the right circumstances, concentric rings propagating outwards from the Galactic center can plausibly produce vertical structures similar to the Triangulum–Andromeda stellar clouds observed in the real Milky Way. On the other hand, using cosmological high-resolution hydrodynamical simulations, \cite{Gomez:2016} found a well defined and strong vertical pattern in the disc of a Milky Way-like galaxy, which was the result of a satellite–host halo–disc interaction and reproduced qualitatively the observable properties of the Monoceros ring.

It is interesting to compare the three-dimensional shape of the corrugation detected in this work to previously known corrugations in the Galactic disc. For instance, \cite{Morganson:2016} mapped the Monoceros Ring in three dimensions using PAN-STARRS1. Specifically, they map an overdensity of stars below the Galactic plane at about $R\sim$14 kpc (see their Fig. 16, lower left panel), and another one above the Galactic plane at about $R\sim$17 kpc (lower right panel in their Fig. 16). The features they detect are not well fitted by a Galactocentric ring, and do not align with the spiral arms. Moreover, the observed features are more distant from the Galactic center in the North than in the South (i.e. leading spiral geometry). These results are in perfect agreement with the properties of the corrugation detected in this work. Given their respective locations in the Galactic disc, and given they share a similar orientation, it is possible that the corrugation presented in this work is an inner ripple with respect to Monoceros ring. We note, however, that previous works (to our knowledge) mapping vertical corrugations in the Galactic disc do not take the warp into account. While this might not be crucial in the regions close to the warp's line-of-nodes (i.e. approximately towards the Galactic anticenter), where the warp amplitude is small, other directions can potentially be biased from the assumption of a perfectly axisymmetric disc, which is clearly not the case of our Milky Way.  

In addition to the vertical structure and motions in the Galactic disc, we also explored the radial component of the individual stellar velocities. The obtained maps indicate that stars in the corrugation are systematically moving outwards ($V_R>0$, see also Figure \ref{fig:vector}, \ref{fig:scatter_Zres_VR_Vzres_young_giants} and \ref{fig:scatter_Zres_VR_Vzres_Cepheids}). It is interesting to note the agreement between the direction of the mean radial motions $V_R$ in the corrugation (pointing toward the outer disc) and the direction of propagation of the wave that would be suggested from the observed oscillatory behaviour $\Delta_Z$-$\Delta_{V_Z}$ if the above wave-like interpretation is assumed.

As mentioned in Section \ref{sec:introduction}, previous studies have suggested evidence for vertical waves in the disc. In particular, \citet{Schoenrich:2018} saw a kinematic signature of the warp as well as an additional wave-like pattern in vertical velocity vs. angular momentum space, with an amplitude of $\sim 1 $ km/s and a scale-length of $\sim 2.5$ kpc in guiding centre radius $R_g$. \citet{Friske:2019} analysed the Gaia DR2 RV sample, covering approximately a volume of $\sim$2-2.5 kpc from the Sun, and found a wave-like pattern in the mean Galactocentric radial velocity vs. $R_g$ space, with a short-wave pattern of order of 1.2 kpc in $R_g$. They found that the wave-peaks in the vertical velocity line up with the outer troughs in the radial velocity wave with a strong correlation. This is in good agreement with the results presented in this work, which are based on both the three-dimensional distribution of stars and their kinematics on a large scale. While the typical age of the tracers is different (our datasets are mostly young), and therefore some differences can be expected, it is possible that our results are connected, or even that they are part of the same oscillatory pattern. In this work we only clearly see about a half oscillation, and the radial extent of the detected pattern suggests a wavelength of at least 4 kpc.

It is generally assumed that vertical waves in the Galactic disc do not have associated radial motions, as radial and vertical motions are usually assumed to be decoupled. The main reasons for this are: 1) The gravitational potential in a galaxy can be approximated as separable into components that affect the radial and vertical motions independently. Vertical motion is influenced by the potential perpendicular to the disc, while radial motion is influenced by the potential within the plane of the disc. 2) Vertical oscillations typically have a higher frequency compared to radial oscillations. In local regions of the galactic disc, stars are assumed to follow approximately epicyclic orbits where vertical oscillations with respect to the disc plane are largely independent and out of phase of their radial motions. 

However, our observational results suggest a different scenario. Our maps suggest that the vertical wave detected here in the Galactic disc also exhibits an associated net radial motion. If this interpretation is correct, we can draw an intriguing analogy with ocean waves, which propagate through water as a result of a circular motion of volumn elements that includes both radial and vertical components, that is, with a component of motion transverse to the direction of wave propagation, and a "longitudinal" component in the direction of wave propagation. These types of waves are referred to as Rayleigh waves. Though in the context of the Galactic disc we are clearly not seeing a surface wave, the coupling of radial and vertical motions here nevertheless suggests that the stars experience both upward and outward movements as the wave propagates.

Our results also indicate that, at least for young stars, the $V_R$ velocity field in the outer disc is dominated by an alternating pattern, whose geometry is correlated with a corrugation in the spatial vertical direction. This is not the first time that an alternate pattern in $V_R$ has been detected. For instance, \citet{GaiaCollKatz:2018, GaiaCollDrimmel:2023}, \citet{Friske:2019,Khanna:2023} and others mapped streaming motions in the Galactic disc, usually with stars typically older than the samples studied  here. \cite{GaiaCollDrimmel:2023} mapped the velocity field of the OB stars, but only reached 1-2 kpc in heliocentric distance due to limited availability of the line-of-sight velocities for these stars. \cite{Eilers:2020} mapped a $V_R$ radial wave using upper Red Giant Branch (RGB) stars, which they interpreted as the signature of the Galactic spiral arms. The radial wave mapped by \cite{Eilers:2020} might be the RGB counterpart of the $V_R$ feature mapped here with young stellar populations. In this work, the $V_R$ feature was discussed in relation to a vertical corrugation, but also compared to spiral arm geometries available in literature, to take into considerations different interpretative scenarios.
Similarly, \cite{Palicio:2023} detected a pattern in the radial action (tracing the absolute value of $V_R$) of disc stars. Recently, \citet{stelea:2024} used a suite of N-body simulations to study the response of the Galactic disc to the Large Magellanic Clouds and the Sagittarius dwarf galaxy (Sgr), and found that the corrugations induced by Sgr can also reproduce the large radial velocity wave seen in the data by \cite{Eilers:2020}. Simulations  \citep[][etc.]{monari2016,Vislosky:2024} have also shown that coupling between gravitational perturbations due to the Galactic bar and spiral density modes \citep[e.g.,][]{LinShu1964,Lin1969}, can generate non-linear effects on the velocity field, and could offer other mechanisms of generating large scale waves in the disc.

Finally, it is important to note that our sample is very young, and thus may be revealing the bulk motions of the gas from which they were born. We therefore cannot expect these young stars to behave as a kinematically relaxed population. Instead they may be illuminating large-scale non-circular motions in the gas disc that could have been excited by a perturbation.

The maps obtained here also imply that the disc of the Milky Way is very active, and exhibit, at the present day, compelling evidence of past (and/or present) perturbations. This has been predicted by numerical simulations \citep{DOnghia:2016,Laporte:2018,BlandHawthorn:2021,TepperGarcia:2022} and has already been shown with real data by a great variety of physical phenomena, like the phase space spirals \citep{Antoja:2018, Binney:2018, Bland-Hawthorn:2019, Laporte:2019, McMillan:2022, Antoja:2023, Alinder:2024}, local vertical waves \citep{Widrow:2012, Widrow:2014, Bennett:2019}, substructures in velocity space and ridges in $V_{\phi}$ vs. $R$ space \citep{Kawata:2018, Antoja:2018, Ramos:2018, Bernet:2022, Antoja:2022, Lucchini:2023}. All these phenomena are showing that the disc is a complex system, and it is not always trivial to link the observational features to their possible generating mechanisms \citep{Cao:2024}.

In this work, we are adding a new element to the (already complicated) picture of the disc of our Galaxy, by presenting the detection of a vertical corrugation on top of the Galactic warp, whose velocities are consistent with a wave propagating towards the outer Milky Way disc. Several theoretical works predicted the presence of similar features in the Galactic disc, triggered by a passing satellite \citep{DOnghia:2016,Laporte:2018,BlandHawthorn:2021,TepperGarcia:2022}. 
However, the specific origin of the feature detected is this work is to date unknown.

The influence of a satellite galaxy has also been proposed by \citet{Thulasidharan:2022} to explain the Radcliffe Wave  \citep[hereafter RW, ][]{Alves:2020}, a coherent structure of dense gas clouds in the Solar Neighbourhood, extending for 2.7 kpc in length. Alternative explanations proposed in the literature include the possibility that the RW is the result from a Kelvin–Helmholtz instability \citep{Fleck:2020}. Moreover, \citet{Swiggum:2022} proposed the scenario that the RW constitutes the gas reservoir of the Local (Orion) Arm, based on the observed offset between those two features. Using the proper motions of young stellar objects anchored inside the clouds, \cite{Li:2022} identified a damped wave-like pattern in the vertical velocities, with a phase difference between the spatial and vertical velocity oscillations of about $2 \pi /3$. Recently, \citet{Konietzka:2024} suggested that the RW is oscillating, with a phase difference of $ \pi /2$, while also drifting away from the Galactic center with a velocity of about 5 km/s. They found that, even though a gravitational interaction with a perturber seems a natural possible origin, the RW's stellar velocities are not fully consistent with models of a perturbed-based scenario. 

It is interesting to compare the RW to the wave detected in this work. Figure \ref{fig:Radcliffe} shows that the two features are located in different regions of the Galactic disc: the RW is a filament located relatively close to the Sun (250 pc at the closest point), while the wave mapped here is more distant (at Galactocentric radii $R \gtrapprox$ 10 kpc, heliocentric distances $d_{HEL} \gtrapprox$ 2 kpc at the closest point) and spans over a much larger area of the outer Galactic disc. Moreover, the mean characteristic wavelength of the RW is about $\sim 2$ kpc \citep{Alves:2020, Konietzka:2024}, while that of the feature mapped here is of at least 4 kpc. Their different properties and locations in the Galactic disc suggest that they should be treated as two distinct features, but it is possible that the two waves are somehow related. For instance, they might be two different fluctuations of a common oscillatory pattern, which permeates the Galactic disc. In this context, it is {\bf worth noting} that the RW has both radial and vertical motions \citep{Konietzka:2024}, with a behaviour similar to the wave studied in this work. Moreover, the RW and the wave mapped here are both based on young tracers. The wave mapped here might be an extension of the RW in the outer disc. However, it is beyond the scope of this paper to assess whether this scenario is realistic or not, and future works will shed further light on whether the two waves are connected or not. 

  
\section{Summary \& conclusions}\label{sec:summary_conclusions}

In this paper, we analyzed the vertical structure and kinematics of the Galactic disc using two samples of young giant stars and Classical Cepheids. The two samples are complementary: Cepheids can be seen to typically large distances ($\sim$ 15 kpc), but are relatively rare ($\sim$ 3000 objects); on the other hand, the young giant sample is mostly confined to within 6-7 kpc in heliocentric distance, but contains significantly more stars ($\sim$ 17000 stars). Notwithstanding their different properties, we found that the two samples give a self-consistent view of the Galactic disc, as expected for two samples with similar young ages. Our main results are summarized in the following.

\paragraph{Warp three-dimensional shape.} The results of our inference based on a classic $m=1$ warp shape are in good agreement with previous results based on young stellar populations. We obtain a warp amplitude of $\sim$ 0.45 kpc at $R \sim$ 12 kpc, and $\sim$ 0.7 kpc at $R \sim$ 14 kpc, in perfect agreement with previous estimates from \citet{Skowron:2019a, Chen:2019, Dehnen:2023,CabreraGadea:2024}. Both samples show that the line-of-nodes begins increasing in the direction of Galactic rotation starting at about R$\gtrapprox$12 kpc, in agreement with previous works \citep{Chen:2019, Dehnen:2023,CabreraGadea:2024}.

\paragraph{A vertical corrugation on top of the warp.} The analysis of the fit residuals $\Delta_Z$ reveals a coherent vertical feature, which is systematically shifted upwards by about 150-200 pc with respect to the best-fit $m=1$ warp shape, located approximately at Galactocentric radiii $R \sim$ 10-14 kpc. Using the sample of young giants, which has the better statistical signal, we find that this feature is not exactly radial: the feature is slightly shifted at different radii depending on the considered azimuthal position (at least, in the region covered by the young giants sample).

\paragraph{The corrugation can be interpreted as a wave propagating outwards.} To study the vertical kinematics, we first infer the warp precession rate using our two samples. We obtain a prograde warp precession \citep[in agreement with][]{Poggio:2020, Cheng:2020, HrannarMcMillan2024}, which is gradually declining with $R$, confirming previous works by \cite{Dehnen:2023} and \cite{CabreraGadea:2024}. After subtracting the warp model in the vertical velocities $V_Z$, we investigate the residuals and find that the corrugation detected in the vertical coordinate $\Delta_Z$ (see point above) has a kinematic self-consistent counterpart. In the same region, but slightly shifted toward the outer disc, we see a similar feature with positive vertical velocities $V_Z$ (after the warp kinematic signature has been subtracted). The shift approximately corresponds to about half of the radial width of the corrugation observed in the Z residuals, so that the lines of the maxima of the vertical velocities roughly coincides with the lines where the Z residuals are zero, indicating an oscillatory behaviour. The comparison of the observed shift with a simple toy model suggests that this is consistent with a wave propagating towards the outer disc.

\paragraph{The stars located in the corrugation exhibit systematic radial motions.} The feature in the outer disc with $\Delta_Z>0$ observed in vertical fit residuals is interestingly appearing in the same position of systematically positive radial motions. Stars that today are located in the positive crest of the corrugation exhibit systematic outwards motions of about 10-15 km/s, while those located outside the crest are moving towards the inner parts of the disc. \break

Taken together, these findings lead us to explore the hypothesis that there is a vertical wave extending over a large portion of the outer disc ($R>10$ kpc) moving away from the Galactic center, possibly involving both vertical and radial motions, that appear to be statistically correlated. 
This wave, detected in young stellar populations, may primarily be in the gaseous component of the Galactic disc, revealed by the kinematics of the young stars which have inherited the bulk motions of the gas from which they were born.

Future works analysing other datasets, as well as different galaxy formation models, will reveal further details on the formation and evolution of the observed feature, as well as a more detailed dynamical description of how waves propagate in the disc of the Milky Way.



\begin{acknowledgements}
EP thanks Dorota Skowron, Joss Bland-Hawthorn, Ralph Schoenrich and Walter Dehnen for useful comments and suggestions. EP also thanks Walter Dehnen for sharing his data. 

This project has received funding from the European Union’s Horizon 2020 research and innovation programme under the Marie Sklodowska-Curie grant agreement N. 101063193. 

RD and EP are supported in part by the Italian Space Agency (ASI) through contract 2018-24-HH.0 and its addendum 2018-24-HH.1-2022 to the National Institute for Astrophysics (INAF).

SK \& RD acknowledge support from the European Union's Horizon 2020 research and innovation program under the GaiaUnlimited project (grant agreement No 101004110). SK acknowledges use of the INAF PLEIADI@IRA computing resources.

This work has made use of data from the European Space Agency (ESA) mission
\gaia\ (\url{https://www.cosmos.esa.int/gaia}), processed by the \gaia\
Data Processing and Analysis Consortium (DPAC,
\url{https://www.cosmos.esa.int/web/gaia/dpac/consortium}). Funding for the DPAC
has been provided by national institutions, in particular the institutions
participating in the \gaia\ Multilateral Agreement.

\end{acknowledgements}


\bibliographystyle{aa} 
\bibliography{mybib} 
   

\begin{appendix}

\section{Additional information on young giants sample selection and distance calculation\label{appendix_sample}}

\begin{figure*}
\centering
\includegraphics[width=18.cm]{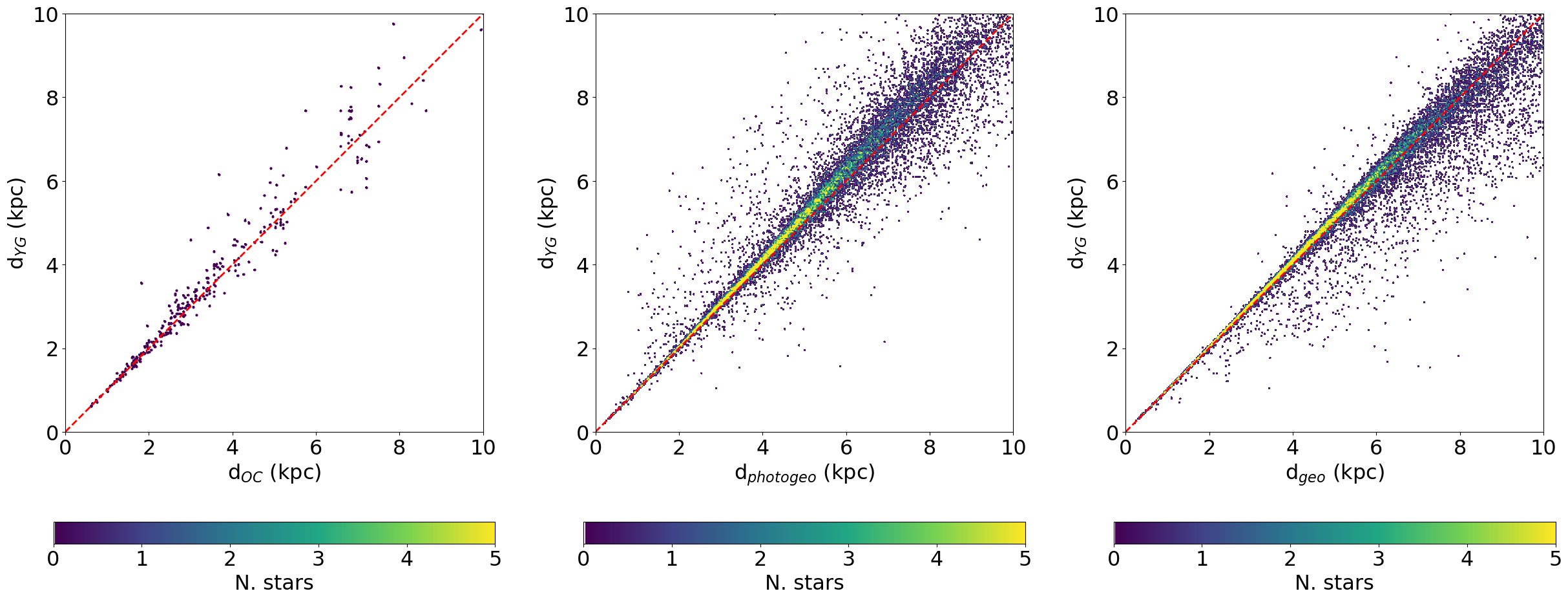}
\caption{ Comparison between different distance estimates. \emph{Left panel:} Subsample of young giants classified as members of open clusters by \cite{Hunt:2023}. The comparison between the distances $d_{YG}$ calculated in Section \ref{subsec:distances} and the distances to each cluster from \cite{Hunt:2023} is shown. The red dashed line shows the line of equality. \emph{Middle panel:} Same as left panel, but now with the distances $d_{YG}$ compared to the photogeo distances from \cite{BailerJones:2021}. \emph{Right panel:} Same as middle panel, but now compared to the geometric distances from \cite{BailerJones:2021}. \label{fig:comparison_dist}}

\end{figure*}

Figure \ref{fig:comparison_dist} shows a comparison between the distances obtained in this work for the young giant sample (see Section \ref{subsec:distances}) and other values taken from the literature. It is well known that heliocentric distances can be studied with great accuracy for open clusters in the Galactic disc \citep[e.g.][]{Cantat-Gaudin:2020, Spina:2022, Cavallo:2024}. For this reason, we cross-matched our sample with the open cluster members from \cite{Hunt:2023}, finding 307 stars in common. In the left panel of Figure \ref{fig:comparison_dist}, we show the comparison between our distances and those given by \cite{Hunt:2023}. The middle and right panels of Figure \ref{fig:comparison_dist} show, respectively, the comparison between the photogeometric and geometric distances from \cite{BailerJones:2021}. 95 \% of our distance estimates agree within 25 \% with the values taken from the literature (for stars within 8 kpc from the Sun).

\begin{figure*}
\centering
\includegraphics[width=18.cm]{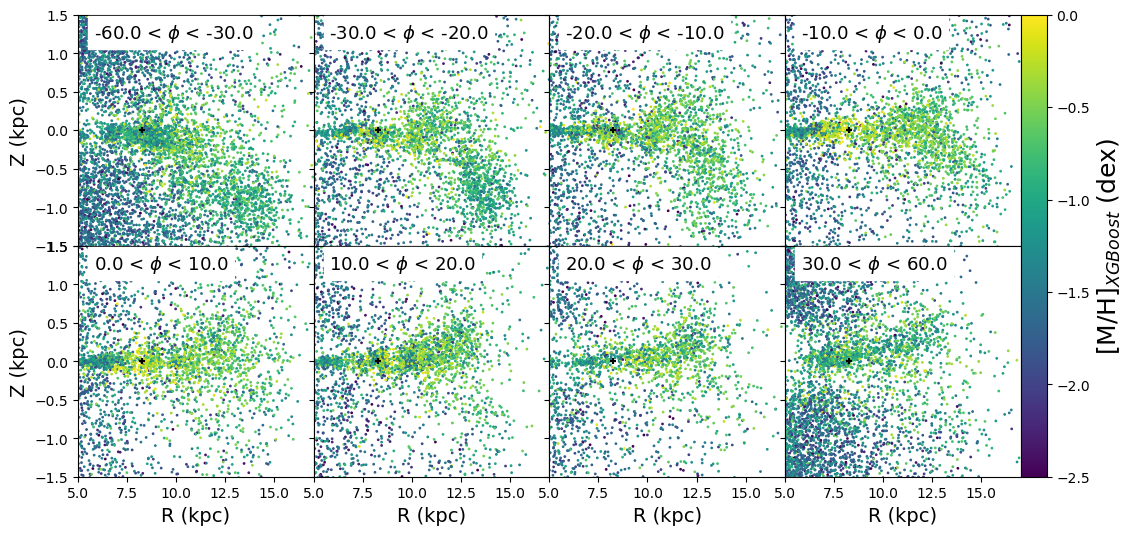}
\caption{Spatial distribution of the young giant candidates before applying the cuts from Section \ref{subsec:contaminants}. Stars are color-coded by metallicity to better highlight the (metal-poor) contaminants at high Z at different slices in Galactic azimuth $\phi$. The location of the Sun is indicated by the black dot.\label{fig:contaminants_RZ_colmet}}

\end{figure*}

Figure \ref{fig:contaminants_RZ_colmet} shows the distribution of our sample in the $R$-$Z$ plane for different slices in Galactic azimuth $\phi$ before applying the cleaning described in Section \ref{subsec:contaminants}. Stars are color-coded by metallicity $[M/H]$ from the XGBoost catalog \cite{Andrae:2023}. As we can see, the selection criteria adopted in Section \ref{subsec:sample_selection} are not enough to select young giants in the Galactic disc: we can see a significant fraction of metal-poor contaminants at large distances from the Galactic plane (which, in contrast, is typically populated by stars that are more metal-rich). 

To further refine our sample from potential contaminants, we look for potential clustered stars in velocity space. We apply the python scikit-learn implementation of the HDBSCAN algorithm \citep{Campello:2013} to our dataset. Based on the output, we find 6 tightly clustered stars in the $V_R$, $V_{\phi}$ and $V_Z$ space, which appear to be located at the same $(l,b)$ coordinates, at the same distance. We compare the $(l,b)$ position of the 6 stars to the globular clusters catalog from \cite{Harris:1996}, and find that they perfectly coincide with the sky coordinates of the globular cluster NGC 6656. Those 6 stars are therefore potential contaminants, and for this reason we removed them from the final catalog.

\section{Sub-selection function}
\label{sec:subsf}
We estimate the sub-selection function using the Gaia query below, run on the archive, in 1 magnitude wide bins in $G$, and at healpix level 5 for bins in $l,b$. The XGBoost catalogue by \cite{Andrae:2023} is based on sources in \gaia{} DR3 for which the XP spectra were published. Furthermore, they restrict to those sources with CatWISE photometry. At the time of writing, the \gaia{} archive only had an official crossmatch provided with AllWISE which is shallower than CatWISE. However, we checked that our catalogue of young giants, owing to their relative brightness, have an entry both for AllWISE \& CatWISE. Thus, we can still proceed with using the ratio-ing method directly on the archive but using the counts based on AllWISE instead, as shown below:

\lstset{language=SQL}
\begin{lstlisting}[caption={\texttt{ADQL} query for the \gaia\ DR3 data considered in the subselection function.},captionpos=b]
SELECT healpix_, phot_g_mean_mag_, COUNT(*) AS n, SUM(selection) AS k 
FROM (SELECT to_integer(GAIA_HEALPIX_INDEX(5,source_id)) AS 
healpix_, to_integer(floor((phot_g_mean_mag - 4)/1.)) 
AS phot_g_mean_mag_, to_integer( IF_THEN_ELSE(has_xp_continuous='true',1.0,0.0) ) 
AS selection FROM gaiadr3.gaia_source as gdr3
left outer join gaiadr3.allwise_best_neighbour as wisenb USING (source_id)
WHERE phot_g_mean_mag > 4 AND phot_g_mean_mag < 15 and wisenb.source_id is not null )
AS subquery GROUP BY healpix_, phot_g_mean_mag_
\end{lstlisting}

\begin{figure}
\centering
\includegraphics[width=0.5\textwidth]{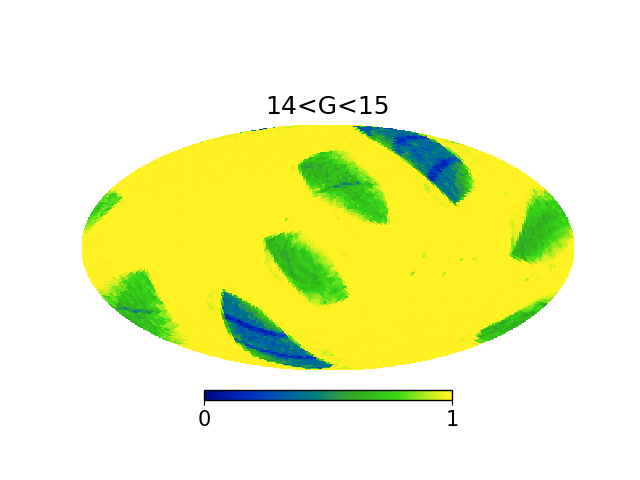}
\caption{Mollweide projection of the selection function $S(l,b,14<G<15)$, for the XGBoost sample in Galactic coordinates. Here we show the magnitude bin $14<G<15$, at HEALpix level 5. The yellow regions indicate the highest completeness, while blue regions are incomplete. \label{fig:sf_14_15}}
\end{figure}

A sky map of the resulting selection function $S(l,b,G)$ for the XGBoost catalogue in the magnitude bin $14<G<15$ is shown in Fig \ref{fig:sf_14_15}. Maps of $S(l,b,G)$ at other apparent magnitudes (not shown) are very similar. 

\section{Corner plots}
\label{sec:corner_plots}

\begin{figure*}
\centering
\includegraphics[width=18.cm]{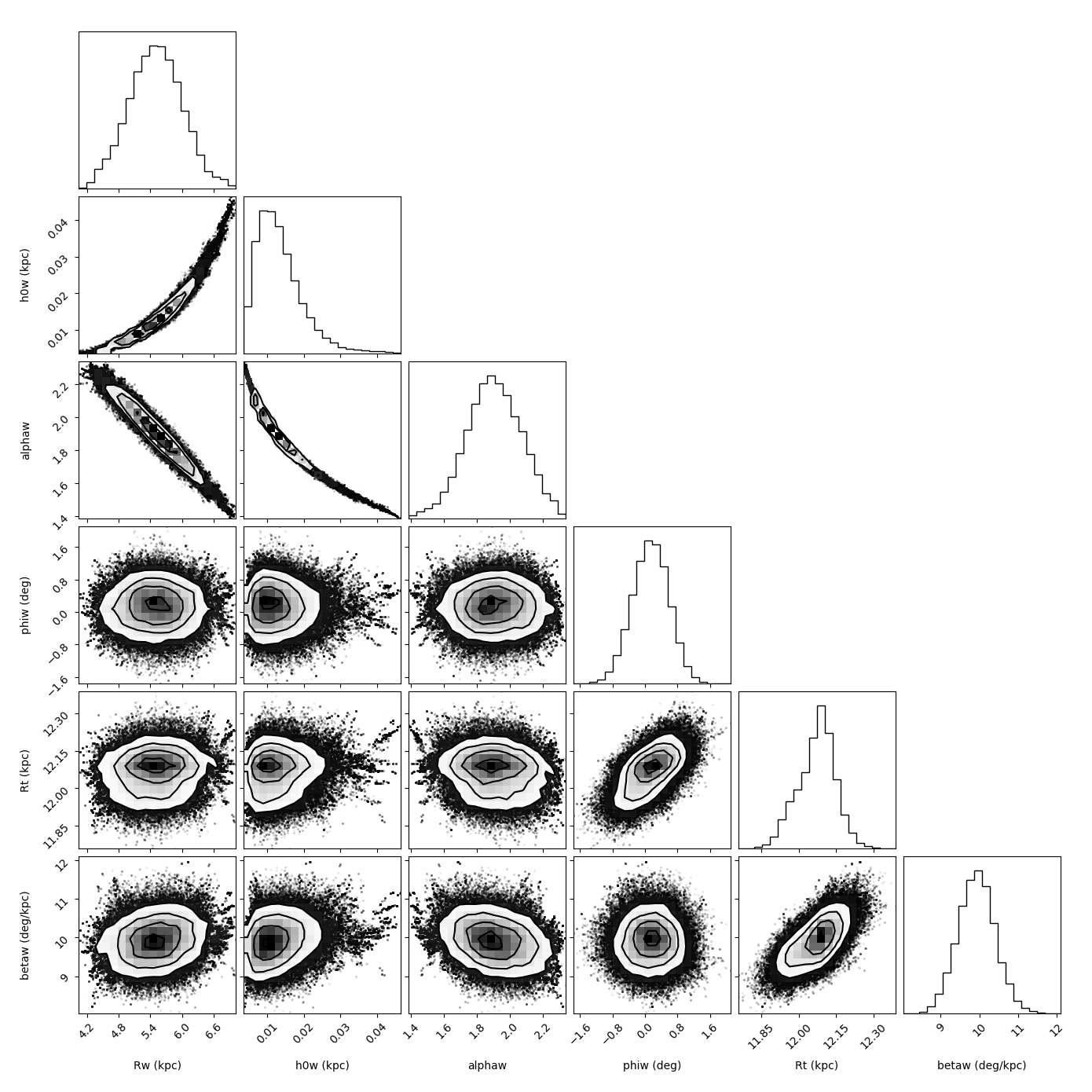}
\caption{Corner plot for the global parametric fit with a twisted line-of-nodes using the young giant catalog. \label{fig:triangle_plot_young_giants}}

\end{figure*}

\begin{figure*}
\centering
\includegraphics[width=18.cm]{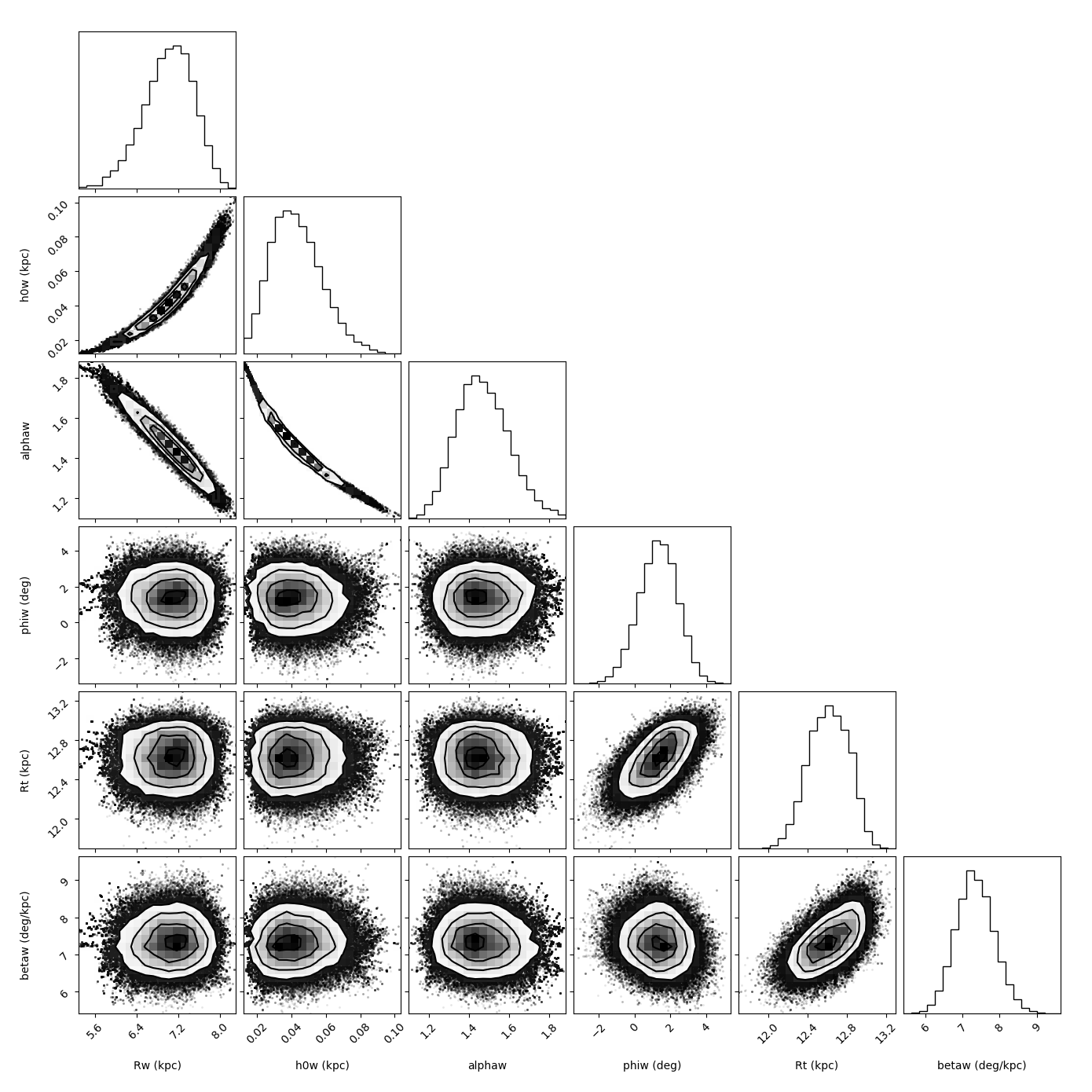}
\caption{Corner plot for the global parametric fit with a twisted line-of-nodes using the Cepheid catalog. \label{fig:triangle_plot_Cepheids}}

\end{figure*}

Figure \ref{fig:triangle_plot_young_giants} and \ref{fig:triangle_plot_Cepheids} show, respectively, the corner plots of the global parametric fit for the young giant sample and the Cepheid sample.

\section{Toy model}
\label{sec:toy_model}

Here we present a simple toy model for a radial corrugation propagating in the Galactic disc, and derive the expected mean vertical velocities $\overline{V_Z}$ based on the $0^{th}$ moment of the collisionless Boltzmann Equation (CBE)\footnote{This corresponds to the first Jeans equation \citep[Eq. 4.204][]{BinneyTremaine:2008}, which differs from the continuity equation only in that it describes conservation of probability rather than that of mass, and replaces the fluid velocity by the mean stellar velocity \citep{BinneyTremaine:2008}.}.

We stress that the goal of this section is not to give a sophisticated and realistic description of the Galaxy, but simply to make a first step towards the understanding of how waves propagate in the disc, and give a possible framework to interpret the observational results presented in this work. We consider a simple double-exponential disc, whose density distribution is modelled by the equation
\begin{equation}
\label{eq:density_disc}
    \rho(R,\phi,z,t)= \rho_0 \,  e^{-\frac{(R-R_{\odot})}{L_R}}
    e^{-\frac{ | z - Z_C(R, \phi,t) |}{h_z}}
\quad,
\end{equation}
where $L_R$ and $h_z$ are the scale-length and scale-height of the disc, respectively, and the coordinate $Z_C(R, \phi, t)$ is the time-dependent vertical displacement of the stellar disc associated with the corrugation. For the sake of brevity, here we analyse the case of a $m=0$ corrugation (i.e. radial ripples), which has no azimuthal dependency. For simplicity, we assume a corrugation described by a sinusoidal behaviour
\begin{equation}
\label{eq:ZC_corrugation}
Z_C(R, t)= A_C \sin{ \left( \frac{2 \pi}{\lambda_C  } (R-R_{C0} - v_C t)
\right)} 
\quad,
\end{equation}
where $A_C$ is the amplitude of the vertical corrugation, $\lambda_C$ is the wavelength of the radial oscillation, $R_{C0}$ is the radial offset with respect to the Galactic center, $t$ is time and $v_C$ is the propagation velocity of the corrugation. For positive (negative) values of $v_C$, the corrugation moves towards the outer (inner) Galactic disc. Moreover, we assume for the moment that the mean radial motions $\overline{V_R}=0$ (we will relax this assumption later) and that the mean vertical velocity $\overline{V_Z}$ does not depend on Z (but only on $(R, \phi)$). These assumptions help us to keep the problem simple, and build upon that by adding new terms, if necessary. Taking the above considerations into account, the CBE in Galactocentric cylindrical Coordinates becomes
\begin{equation}
\label{eq:cbe_zeroth_galcoords_simplified}
    \frac{ \partial \rho}{\partial t} 
    + \overline{V_Z} \frac{ \partial \rho}{\partial Z}
    =0
\quad,
\end{equation}
and we can solve it for the vertical velocity
\begin{equation}
\label{eq:vz_simple_case}
    \overline{V_Z} = - 2 \pi \frac{A_C}{\lambda_C} \, v_c \, \cos{ \Bigl( \frac{2 \pi}{\lambda_C} (R-R_{CO} ) \Bigr)}
\quad,
\end{equation}
which is evaluated at $t=0$ (today). Figure \ref{fig:z_vz_1D} shows the displacement in Z (blue curve) and in $\overline{V_Z}$ (red curve) caused by a vertical corrugation $Z_C$ with $A_C=0.2$ kpc, $\lambda_c=4$ kpc, $R_{C0} = 10$ kpc, and propagating toward the outer parts of the disc with a velocity $v_c$=10 km/s (left panel), or towards the inner parts of the disc at $v_c$=-10 km/s (right panel). 
 
As we can see, in this simple experiment, the vertical velocity pattern is slightly shifted outwards (inwards) with respect to the vertical displacement for a corrugation moving towards the outer (inner) parts of the disc. Similar considerations remain valid when less simplifying assumptions are made to Equation \ref{eq:cbe_zeroth_galcoords_simplified}, for instance including a non-zero $\overline{V_R}$ pattern similar to the one observed in the data, or an azimuthal dependency in the shape of the corrugation (e.g. assuming a pitch angle of $\pm 10{\degree}$).

\end{appendix}

\end{document}